\documentclass[sigconf]{acmart}

\usepackage{textcomp}
\usepackage{url}
\usepackage{multirow}
\usepackage{subcaption}
\usepackage[linesnumbered,ruled,vlined]{algorithm2e}

\AtBeginDocument{%
  \providecommand\BibTeX{{%
    \normalfont B\kern-0.5em{\scshape i\kern-0.25em b}\kern-0.8em\TeX}}}

\copyrightyear{2023} 
\acmYear{2023} 
\setcopyright{acmlicensed}\acmConference[WWW '23]{Proceedings of the ACM Web Conference 2023}{April 30-May 4, 2023}{Austin, TX, USA}
\acmBooktitle{Proceedings of the ACM Web Conference 2023 (WWW '23), April 30-May 4, 2023, Austin, TX, USA}
\acmPrice{15.00}
\acmDOI{10.1145/3543507.3583274}
\acmISBN{978-1-4503-9416-1/23/04}

\begin{document}

\newcommand\mycommfont[1]{\footnotesize\ttfamily\textcolor{blue}{#1}}

\newcommand{\ourmethod}{\texttt{CausIL}}
\newcommand{\etal}{\textit{et al.}}

\newcommand\independent{\protect\mathpalette{\protect\independenT}{\perp}}
\def\independenT#1#2{\mathrel{\rlap{$#1#2$}\mkern2mu{#1#2}}}

\title{\textit{CausIL}: Causal Graph for Instance Level Microservice Data}


\author{Sarthak Chakraborty}
\affiliation{%
  \institution{Adobe Research}
  \city{Bangalore}
  \country{India}}
\email{sarchakr@adobe.com}

\author{Shaddy Garg}
\authornote{Equal Contribution}
\affiliation{%
  \institution{Adobe}
  \city{Bangalore}
  \country{India}}
\email{shadgarg@adobe.com}

\author{Shubham Agarwal}
\authornotemark[1]
\affiliation{%
  \institution{Adobe Research}
  \city{Bangalore}
  \country{India}}
\email{shagarw@adobe.com}

\author{Ayush Chauhan}
\authornote{Work done at Adobe Research, India}
\affiliation{%
  \institution{The University of Texas at Austin}
  \city{Austin}
  \country{USA}}
\email{ayushchauhansma.97@gmail.com}

\author{Shiv Kumar Saini}
\affiliation{%
  \institution{Adobe Research}
  \city{Bangalore}
  \country{India}}
\email{shsaini@adobe.com}

\renewcommand{\shortauthors}{Chakraborty, et al.}

\begin{abstract}

AI-based monitoring has become crucial for cloud-based services due to its scale. A common approach to AI-based monitoring is to detect causal relationships among service components and build a causal graph. Availability of domain information makes cloud systems even better suited for such causal detection approaches. In modern cloud systems, however, auto-scalers dynamically change the number of microservice instances, and a load-balancer manages the load on each instance. This poses a challenge for off-the-shelf causal structure detection techniques as they neither incorporate the system architectural domain information nor provide a way to model distributed compute across varying numbers of service instances. To address this, we develop \ourmethod{}, which detects a causal structure among service metrics by considering compute distributed across dynamic instances and incorporating domain knowledge derived from system architecture. Towards the application in cloud systems, \ourmethod{} estimates a causal graph using instance-specific variations in performance metrics, modeling multiple instances of a service as independent, conditional on system assumptions. Simulation study shows the efficacy of \ourmethod{} over baselines by improving graph estimation accuracy by $\sim$25\% as measured by Structural Hamming Distance whereas the real-world dataset demonstrates \ourmethod{}'s applicability in deployment settings.
\end{abstract}



\maketitle

\section{Introduction}

Modern cloud-based applications follow a microservice architecture \cite{balalaie2016microservices}, consisting of a large number of components connected through complex dependencies~\cite{netflix, uber} that run in a distributed environment. These applications have a simple development process and flexible deployment in general~\cite{newman2021building}. 
A modern microservice ecosystem~\cite{nadareishvili2016microservice} deploys multiple instances of the same service (which are sometimes referred to as pods~\cite{pods}, though similar in meaning) and often increases and decreases their number in response to the change in the load and utilization. The number of unique instances is numerous, short-lived and varying over time. As an illustrative example, in a small microservice at Adobe, 1117 unique instances were spawned within 3 months, with an average of 8 instances active at any time instant. The median life of the instances was only 350 mins. Auto-scalers are configured to automatically scale the number of instances up or down depending on the load and utilization, and a load balancer manages and distributes the load to each of these instances. Due to their large scale and complexity, such architectures are vulnerable to failures. Hence, to maintain the availability and reliability of the system, an accurate structural understanding of the system is required to perform multiple performance diagnosis tasks~\cite{chen2014causeinfer, meng2020localizing, liu2021microhecl, laguna2013automatic, nguyen2011pal, nguyen2013fchain, chen2019outage}.

Building a causal dependency graph to represent system architecture is an approach being used extensively~\cite{chen2014causeinfer, chen2019outage, liu2021microhecl, wu2021microdiag, ikram2022root} in the domain of performance diagnosis. Each deployed instance of a microservice is monitored by numerous system indicator metrics~\cite{netflixblog2015} like load request, resource utilization, HTTP errors, etc., thus resulting in thousands of such metrics for an entire system. The objective is to learn a graph with causal dependencies among all these metrics using \textit{causal structure estimation} approaches, such that metrics form the nodes and edge $(a,b)$ denotes that metric $a$ causally affects metric $b$. A failure in one metric can be traced back to the subsequent anomalies in other metrics across multiple microservices using the discovered causal graph.


With multiple instances getting deployed and remaining active for a microservice, it is natural to use metrics from each of the instances in the causal structure estimation. Aggregating metric values over all instances reduce the instance-specific variations of metrics, hence losing a significant amount of information~\cite{moghimi2020information}, which even we show in \S \ref{sec:averaging} and \S \ref{sec:results}. However, none of the off-the-shelf causal discovery algorithms~\cite{spirtes2000causation, ramsey2017million, liu2021microhecl} can handle the task of instance-level causal structure learning. Past works~\cite{chen2014causeinfer, gan2021sage, li2022causal, meng2020localizing} build a causal graph only at an aggregate level and overlook the deployment strategy of microservices over multiple instances. However, even a few instances failing can degrade the quality of service, which might not be captured in an aggregate statistics. Hence, one must perform a diagnosis at the instance level, which can only be possible by capturing instance-specific variations in a causal graph. However, with each instance being short-lived and the total number of instances varying over time, modeling at instance-level is a challenging task, which we aim to solve. 


In this work,  we propose a novel instance-specific causal structure estimation algorithm, which to the best of our knowledge is the first-of-its-kind, that considers metric variations for each instance deployed per service while building the causal structure, implicitly modeling the decisions of auto-scalers and load balancers. It further incorporates domain knowledge backed by system-based metric semantics combined with intuitive assumptions in a scalable way to improve the accuracy of the structure detection algorithm. We call our approach Causal Structure Detection using Instance Level Simulation \(\ourmethod{}\). We have made
our code publicly available\footnote{\href{https://github.com/sarthak-chakraborty/CausIL}{https://github.com/sarthak-chakraborty/CausIL}}. We validate our approach on synthetic and semi-synthetic datasets, on which it outperforms the baselines by $\sim25\%$. We also observe that a large impact is seen by incorporating domain knowledge where accuracy improves by $3.5\times$, as measured by Structural Hamming Distance. We further validate \ourmethod{} on a real-world dataset and show performance gain against baselines. The contributions of our work can be summarized as:
\begin{enumerate}
    \item We study multiple ways of incorporating instance-specific variations in metrics including aggregation strategies and discuss their shortcomings empirically
    \item We propose a novel causal structural detection algorithm at the instance level (\ourmethod{}) to model instance-level data for microservices. It accounts for multiple data points belonging to multiple instances at each time period due to distributed compute (load-balancer) as well as can model the dynamic number of instances (spawned by an autoscaler), wherein each instance can be short-lived.
    \item Domain Knowledge Inclusion: Based on practical assumptions derived from the service dependency graph and metric semantics, we provide a list of general rules in the form of prohibited edges along with ways to incorporate them in \ourmethod{} to improve computation time and accuracy.
\end{enumerate}



\section{Related Work} \label{sec:related-work}

With the recent upsurge of work in the field of performance diagnosis of large and complex modern microservice architectures~\cite{chen2014causeinfer, meng2020localizing, mondal2021scheduling, brandon2020graph, ikram2022root}, researchers at many cloud-based companies are actively working on alerting and monitoring solutions ~\cite{chen2019outage, wang2019grano, li2020gandalf}. Alerting services like \textit{Watchdog}, \textit{New Relic} and \textit{Splunk}  diagnose systems by constructing causal graphs between services or using thresholding techniques. These service~\cite{new-relic} monitor performance for each instance by setting alerts for each instance deployed for a microservice. As a result, service reliability tools that use causal dependency graphs should use instance-level data rather than aggregated data for accurate representation. This also allows for the identification and isolation of issues that would otherwise go unnoticed with aggregated data.


Initial works~\cite{nguyen2013fchain, nguyen2011pal} have built dependency graphs among services (also called as service call graphs) to diagnose performance during faults, which show control flow from one service to another, which allow analysis of faults at the service level granularity, that is identifying which service is faulty. Service dependency graphs are also used to answer ``what-if" based system questions on bandwidth management and application latencies~\cite{tariq2008answering, tariq2013answering,jiang2016webperf,hu2018log2sim}. Grano~\cite{wang2019grano} builds a causal dependency graph among physical resources for fault diagnosis. However, performance diagnosis is often demanded at a more granular level.


Multiple works~\cite{chen2014causeinfer, meng2020localizing, wu2021microdiag, wang2018cloudranger, wang2021groot} aim to build a causal graph at the performance metric level to diagnose a system in terms of faulty metrics of a service. They utilize the PC algorithm ~\cite{spirtes2000causation}, a causal structure estimation algorithm, to construct a causal dependency graph at the performance metric level, with each metric representing a node. \cite{meng2020localizing} proposed a variation of the PC algorithm  to build the causal graph by utilizing the temporal relationships between the metrics of various services. Other nuances in causal graph construction involves using knowledge graph in conjunction with PC algorithm~\cite{qiu2020causality}, or using alerts which are triggered when a metric crosses a threshold~\cite{chen2019outage}. However, these approaches are agnostic to the deployment strategy of a microservice which involves multiple instances being spawned and auto-scaled for load-balancing.During a high-load period, aggregating information across all deployed instances to construct the causal graph can result in information loss.~\cite{moghimi2020information}. Understanding instance-specific metric variations is critical for better capturing system state. 

Furthermore, incorporating knowledge of the system architecture can improve the accuracy of the estimated causal graph by removing unnecessary or redundant connections between metrics and enforcing connections that are inherent in microservice systems. Some works~\cite{gan2021sage, li2022causal} have developed a causal Bayesian network of the system using system knowledge and causal assumptions. However, to the best of our knowledge, no previous studies have combined instance-level variations in metric data with system knowledge to estimate a causal graph at the performance metric level, which is the main contribution of our research. 

\section{Problem Formulation} \label{sec:problem-statement}

\subsection{Preliminaries}
\begin{figure}[h]
    \centering
    \includegraphics[width=0.9\columnwidth]{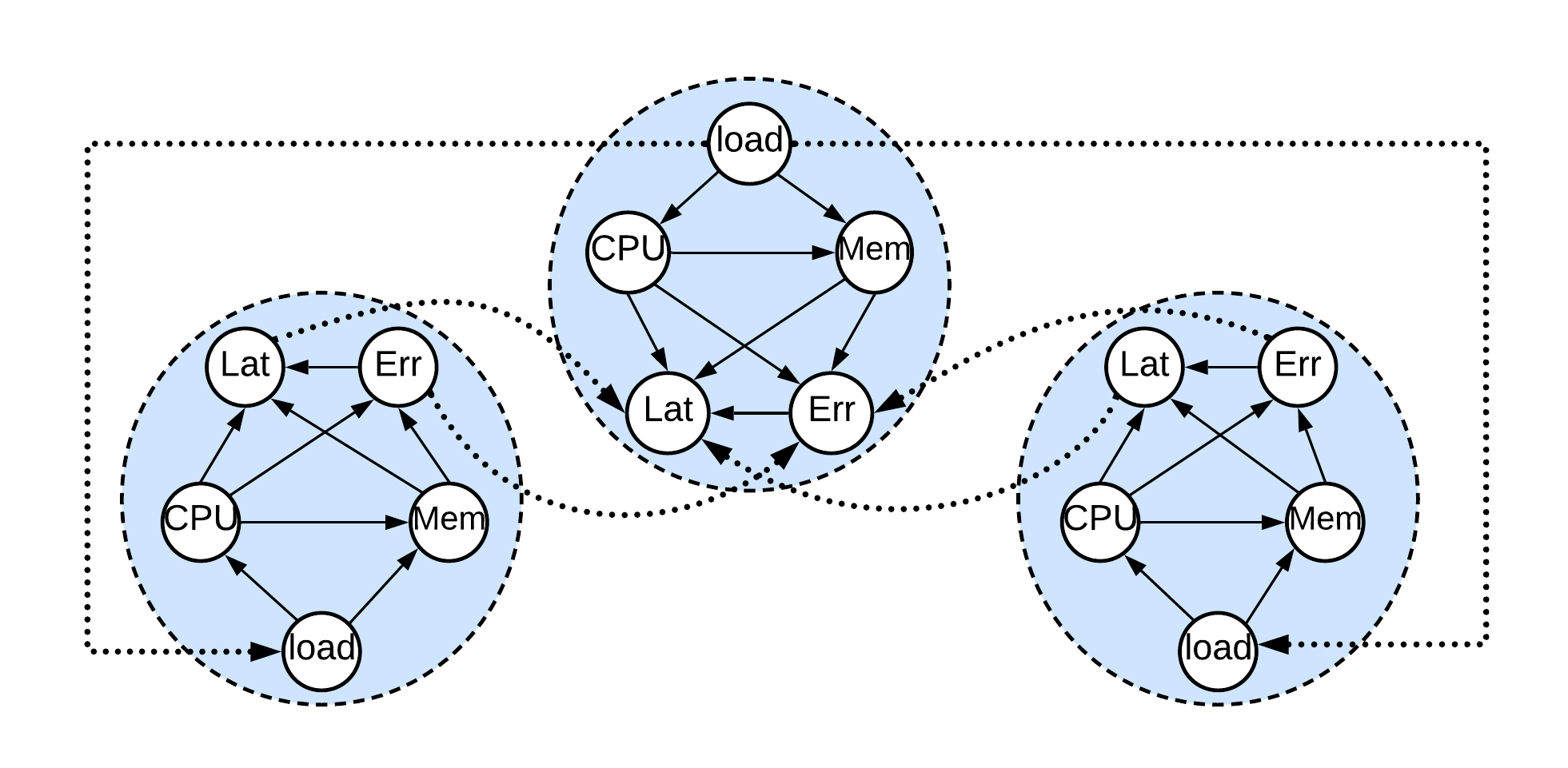}
    \Description[Causal Dependency Graph among performance metrics]{A causal graph between performance metrics of multiple services  where workload, error and latency metrics are connected across services.}
    \caption{\textit{Causal dependency graph between performance metrics of 3 services. Each service is denoted by a dotted blue circle. Dotted lines indicate dependencies that span across services. The white solid circles are the performance metrics}.}
    \label{fig:causal_graph}
    \vspace{-0.4cm}
\end{figure}

A causal structure for a microservice based application (also called dependency graph) is built either with nodes as services or as various performance metrics like latency, load request, utilization etc. The latter approach provides a richer understanding of the entire system and fault localization can be achieved at a more granular level~\cite{chen2014causeinfer, wang2018cloudranger, meng2020localizing, ikram2022root} by indicating that a particular metric of a service is faulty. Our solution utilizes the second approach, with a causal graph at the metric level (Fig. \ref{fig:causal_graph}) where nodes will indicate a set of performance metrics (denoted as parent set for a metric) being responsible for causally influencing an affected metric. Throughout this paper, we have interchangeably used the term service and microservice.


\subsubsection{Causal Graph Structure}
Let $\mathcal{G}(V,E)$ be a causal graph (directed acyclic graph DAG) where the set of nodes $V$ are the metrics  observed in the system, and $E$ are the causal edges between those metrics. An edge $v_i \rightarrow v_j$, where $v_i, v_j \in V$ in $\mathcal{G}$ denote that metric $v_i$ is the cause for the metric $v_j$ and $v_j \independent V' | v_i$, where $V' = V - {v_i}$, that is metric $v_j$ is conditionally independent of all other metric given $v_i$. The process to estimate a causal graph that is faithful~\cite{pearl2000models} to a given dataset is known as \textit{causal structure estimation/discovery}.


\subsection{Problem Definition}
For a microservice $\mathcal{S}$, let $x_{ijt}$ be the value for the $i^{th}$ metric (e.g., latency, CPU utilization, etc.) of $j^{th}$ instance of $\mathcal{S}$ at $t^{th}$ time period. We suppress the subscript for $\mathcal{S}$ for ease of exposition. Let $x_{ijt}$ be a child metric causally dependent on the set $\mathcal{P}(x_{ijt})$ of parent metrics. Thus the conditional distribution of $x_{ijt}$ given $\mathcal{P}(x_{ijt})$ can be written as:
\begin{equation} \label{eq:condprob}
    x_{ijt} = f_{ij}(\mathcal{P}(x_{ijt})) + \epsilon_{ijt}
\end{equation}
where, $\epsilon_{ijt}$ is the residual. The task of a causal structure estimation algorithm is to identify the parents $\mathcal{P}$ for each metric $x_{ijt}$ as well as the causal function $f_{ij}(.)$. Given the set of causal parents for each child metric, an estimator $\hat f_{ij}$ estimates the strength of the relationship between the parent metrics and the child metric. \ourmethod{} uses fGES~\cite{ramsey2017million}, a score-based causal discovery algorithm, which is a fast and parallelized form of Greedy Equivalent Search (GES)~\cite{meek1997graphical, chickering2002learning} designed for discovering DAGs with random variables denoting the causal structure. Following literature, we use a penalized version of the Bayesian Information Criterion (BIC)~\cite{schwarz1978estimating} as the scoring function which is maximized to select the appropriate causal structure for continuous variables. Modeling performance metrics of multiple instances for a service in a causal graph $\mathcal{G}$ is not trivial, and no principled strategy exists in literature, which we circumvent. 


\section{Solution Overview}
In this section, we first describe the causal assumptions that we make while determining the parents of a particular metric node in the causal graph, followed by two preliminary approaches and their shortcomings in modeling a microservice system. We then propose a novel method named \ourmethod{} that leverages metrics for each instance spawned for a microservice while estimating the causal structure of the entire system.
 
\subsection{Metrics Data and Causal Assumptions} \label{sec:data_gen}

Multiple performance metrics are observed in a microservice system, and there is a little global knowledge of which metrics are more important. Previous approaches have used various feature selection methods~\cite{Thalheim2017Sieve} to select a subset of \textit{influential} metrics. However, such approaches are out of the scope for this paper. For each microservice, \ourmethod{} observes performance metrics in 5 different categories - (i) Workload ($W$), (ii) CPU Utilization ($U^c$), (iii) Memory Utilization ($U^m$) (iv) Latency ($L$), and (v) Errors ($E$), which are widely used for system monitoring in industries and are termed as the golden signals~\cite{googlesrebook}. Any metric can be essentially classified into one of these categories. For example, the latency metric encapsulates disk I/O latency, web transaction time, network latency, etc. Throughout this paper, we consider these broad categories of metrics in our formulation, while individual monitoring metrics can be plugged into the categories.

\begin{figure}[!ht]
    \centering
    \includegraphics[width=0.4\linewidth]{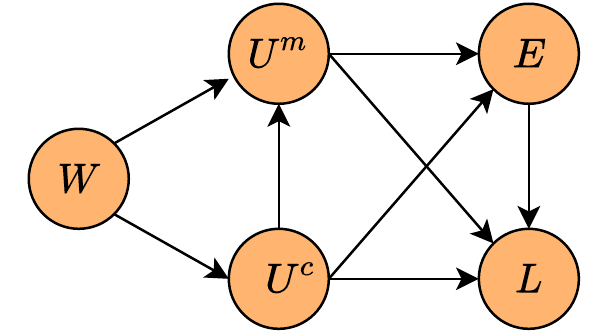}
    \Description[Causal Dependency Graph of Performance Metrics for a Service Instance]{This graph depicts the causal relationship between various performance metrics for an instance of a service. The root node represents the workload, and the leaf node represents latency.}
    \caption{\textit{Causal metric graph for each instance of a service.}}
    \vspace{-0.3cm}
    \label{fig:metric_graph}
\end{figure}

Similar to a previous approach~\cite{li2022causal}, we define certain causal assumptions between the metric categories based on domain knowledge of system engineers to define a causal metric graph (Fig. \ref{fig:metric_graph}). A \textit{workload} request at a microservice in turn demands resource \textit{utilization} within the microservice, while the \textit{latency} is the final effect of the request and hence is the leaf node in the metric graph. Such an assumption holds for all the microservices and their instances. 

For request delivery across microservices, we employ the assumption that holds for a generic microservice architecture, that is, \textit{the latency and error metrics of a service depend on the callee microservice, while the workload depends on the workload of the caller microservices.} Concisely, for a request trace from microservice A to B, where B is the callee microservice and A is the caller microservice, the  ground-truth interservice edges that we consider in our formulation are $W^A_t \rightarrow W^B_t$, $L^B_t \rightarrow L^A_t$, and $E^B_t \rightarrow E^A_t$. A causal metric graph with a different selection of metrics can be easily integrated through data-specific domain knowledge.

\subsection{Preliminary Approaches}

We describe two approaches and identify how they fail to capture the system state in the estimated dependency graph. 

\subsubsection{Instances as Dedicated Nodes} \label{sec:individual-node}
 A na\"ive approach to model each instance of a service in the causal graph $\mathcal{G}$ is to have a dedicated node for each $x_{ij}$ representing a metric from each instance of a service $\mathcal{S}$. However, whether causal functions are distinct for each instance of $\mathcal{S}$ can be tested by running the hypothesis $f_{ij} = f_i$ $\forall j$, given enough data. But, due to the short-lived and ever-changing nature of instances being spawned, some instances might get killed and not be re-spawned again with identical physical characteristics. Hence, available metric data for some instances might be scarce and a stable relational function might not be obtained, making the solution infeasible. In addition, with the number of instances varying over time due to auto-scaler decisions, a dynamic causal graph might be needed if each node in the causal graph corresponds to an individual instance of a service. Moreover, the na\"ive solution incurs huge computation to run causal structure estimation with 1000's of instances deployed per service in a large microservice ecosystem.


\subsubsection{Value Aggregation across Instances} \label{sec:averaging}

For each microservice $\mathcal{S}$, this approach (termed as Avg-fGES) groups relevant metrics that causally affect metrics in $\mathcal{S}$ to identify the causal structure of $\mathcal{S}$, i.e., all the five metrics for each instance of $\mathcal{S}$, along with the workload for all instances of caller services and latency and error metrics for all instances of callee services of $\mathcal{S}$ in the service call graph. However, the number of metric values for each service (which is equal to the number of instances) changes with time due to the changing number of instances. Hence at each time instant $t$, to consolidate the metric values for all the instances of a service $x_{ijt} \forall j$, Avg-fGES averages (other aggregation methods can be used) them across the number of instances to form a single aggregated metric $y_{it}$. Formally,
\begin{equation}
    y_{it} = \frac{1}{num\_inst(\mathcal{S})} \sum_{j=1}^{num\_inst(\mathcal{S})} x_{ijt}
\end{equation}

Thus, using the aggregated metric values $y_{it} \forall i$ filtered for service $\mathcal{S}$ and running fGES yields the causal structure.

\begin{figure}[t]
    \centering
    \includegraphics[width=0.35\textwidth]{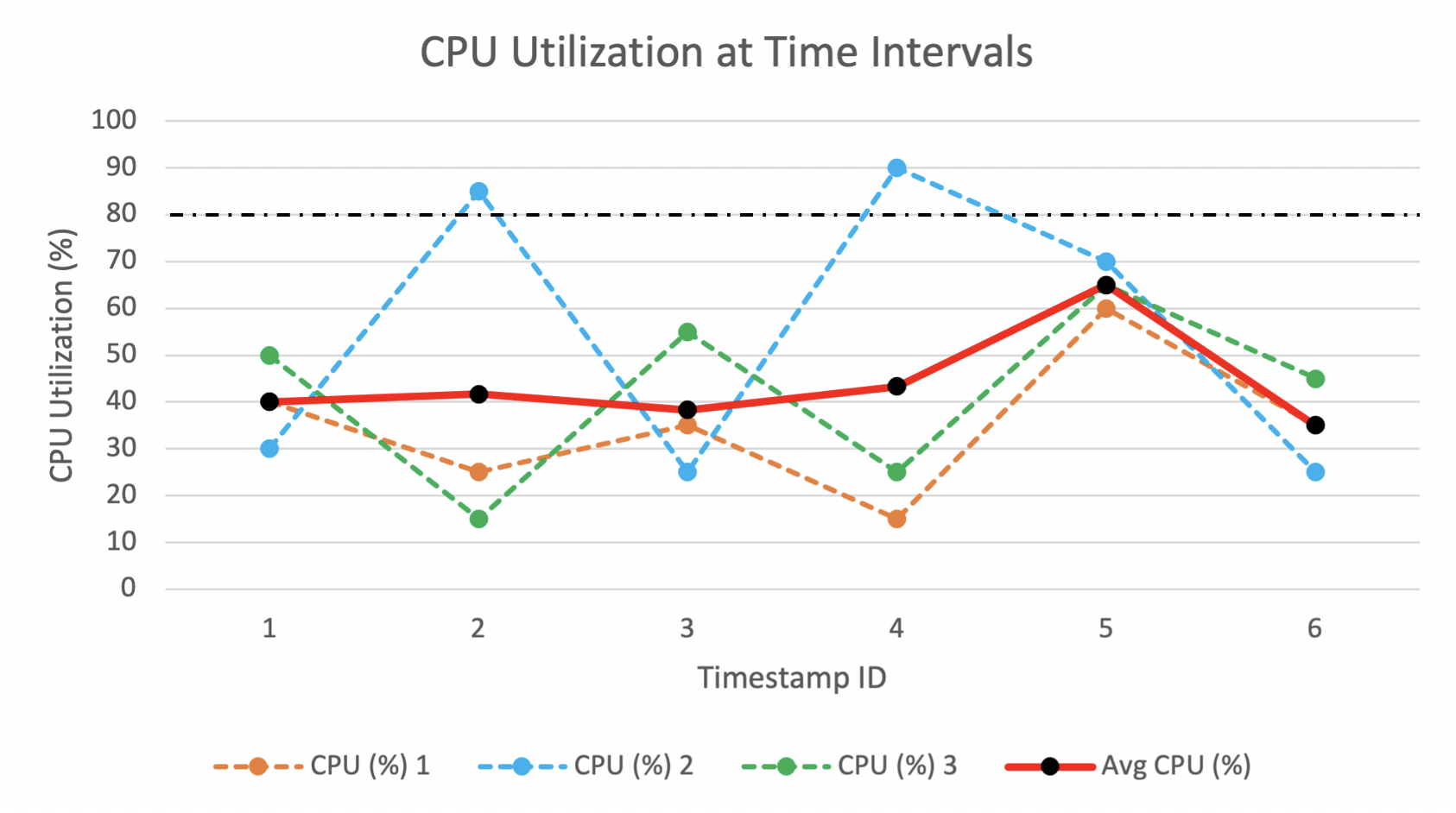}
    \Description[Line chart showing variation of CPU Utilization]{This chart depicts the variation of CPU Utilization for three instances and their average against timestamp and shows how averaging the metric values across all instances dilute the effect of the value of certain instances. Although CPU utilization of one instance is above a threshold, the average over 3 instances is below the threshold.}
    \caption{\textit{Variation of CPU Utilization against timestamp for 3 instances and their average CPU utilization.}}
    \vspace{-0.3cm}
    \label{fig:avg-latency}
\end{figure}

\textbf{Why Aggregation Fails?} Averaging the metric values across all instances for a service results in loss of information~\cite{moghimi2020information, marvasti2010quantifying} and hence fails to capture the true dependency relationships among the metrics. A subset of instances of a service might suffer from high utilization periods, but averaging the metric values across all instances dilute the effect of certain instances (Fig. \ref{fig:avg-latency}), and we lose critical dependencies for such instance exhibiting extreme behavior due to aggregation. Hence the causal relationships will not capture such extreme behaviors.

\begin{figure}[t]
\centering
    \begin{subfigure}[b]{0.49\columnwidth}
        \includegraphics[width=\textwidth]{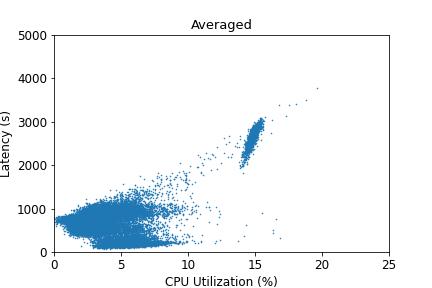}
        \caption{}
    \end{subfigure}
    \begin{subfigure}[b]{0.49\columnwidth}
        \includegraphics[width=\textwidth]{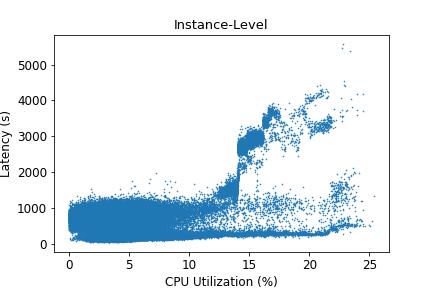}
        \caption{}
        \label{fig:inst_vs_avg_inst}
    \end{subfigure}
\Description[Scatter plot showing the relationship
between latency and cpu utilization for average vs instance level data]{In (a), average metric value for is plotted which shows distortion. In (b) metric value for each instance is plotted, and it shows a quadratic-type relationship between latency and cpu utilization. It shows how averaging data across instances changes the relationship between two metrics.}
\caption{\textit{Averaging data across instances changes the relationship between two metrics. In (b), where metric value for each instance is plot, it shows a quadratic-type relationship between latency and cpu utilization, which gets distorted in (a) due to averaging.}}
\vspace{-0.3cm}
\label{fig:inst_vs_avg}
\end{figure}

Furthermore, system metrics exhibit non-linear dependencies amongst themselves~\cite{masouros2020rusty}, which get deformed when metric values are averaged (Fig. \ref{fig:inst_vs_avg}). For example, as shown in Fig. \ref{fig:inst_vs_avg_inst} obtained from real data, latency shows a non-linear dependency with CPU utilization, with dependencies differing between high and low utilization periods. Consequently, \ourmethod{} models non-linear dependencies among the performance metrics at an instance level. Similar issues prevail for any aggregation metric (sum, max, percentile, etc.). Summing-up metrics like CPU-Usage might drown out signals from faulty instances on receiving noise from normal instances. For example, the CPU utilization of two instances increasing by 10\% might be normal but one increasing by 20\% while another staying same might be a fault signal, the difference not being captured by just summing up the utilization values.

\subsection{\ourmethod{}: Proposed Approach} \label{sec:approach}

To alleviate the issues elicited in the above sections, we propose \ourmethod{} that models data observed in multiple instances for each service. We further describe how system domain knowledge can be used to improve accuracy of any causal graph learning algorithm.

\subsubsection{Instances as IID Draws}
To motivate the design of \ourmethod{}, we look at how Kubernetes or similar platforms deploy services where multiple instances (or pods) of a microservice are launched with the same configurations~\cite{k8horizontal}. Secondly, these containerized instances are isolated and mostly operate independently of each other~\cite{ibm-container}, making them conditionally independent on upstream and downstream metrics like utilization and latency. Their deployment in general, guarantees that there is no interference from other instances of the same service due to co-location~\cite{googleborg}. It can be safely inferred that conditioned on the load received at the service, its different instances are essentially independent and identical, which we use as an explicit assumption in modeling \ourmethod{}. In short, let $W_{j}, U_j$ be the workload and CPU Usage for instance $j$ of a service respectively, then the conditional distribution $U_j | W_j$ is independent of $U_k | W_k$, whereas whereas $U_j$ and $U_k$ are almost certainly dependent.

During high-load requests, the auto-scalers scale up the number of instances for a particular service, and the load balancer distributes the load almost equally (as we observe from real data) to each of the instances. Performance metrics for an instance depend on the amount of load distributed to only that instance. Following our assumption of instances being conditionally independent given the workload, it implies that the causal function for each instance is the same. Hence, Eq. \ref{eq:condprob} can be rewritten as:


\begin{equation} \label{eq:condprobpool}
    x_{ijt} = f_{i}(\mathcal{P}(x_{ijt})) + \epsilon_{ijt}
\end{equation}

\begin{figure}[t]
   \centering
     \begin{subfigure}[b]{0.54\columnwidth}
         \centering
         \includegraphics[width=\columnwidth]{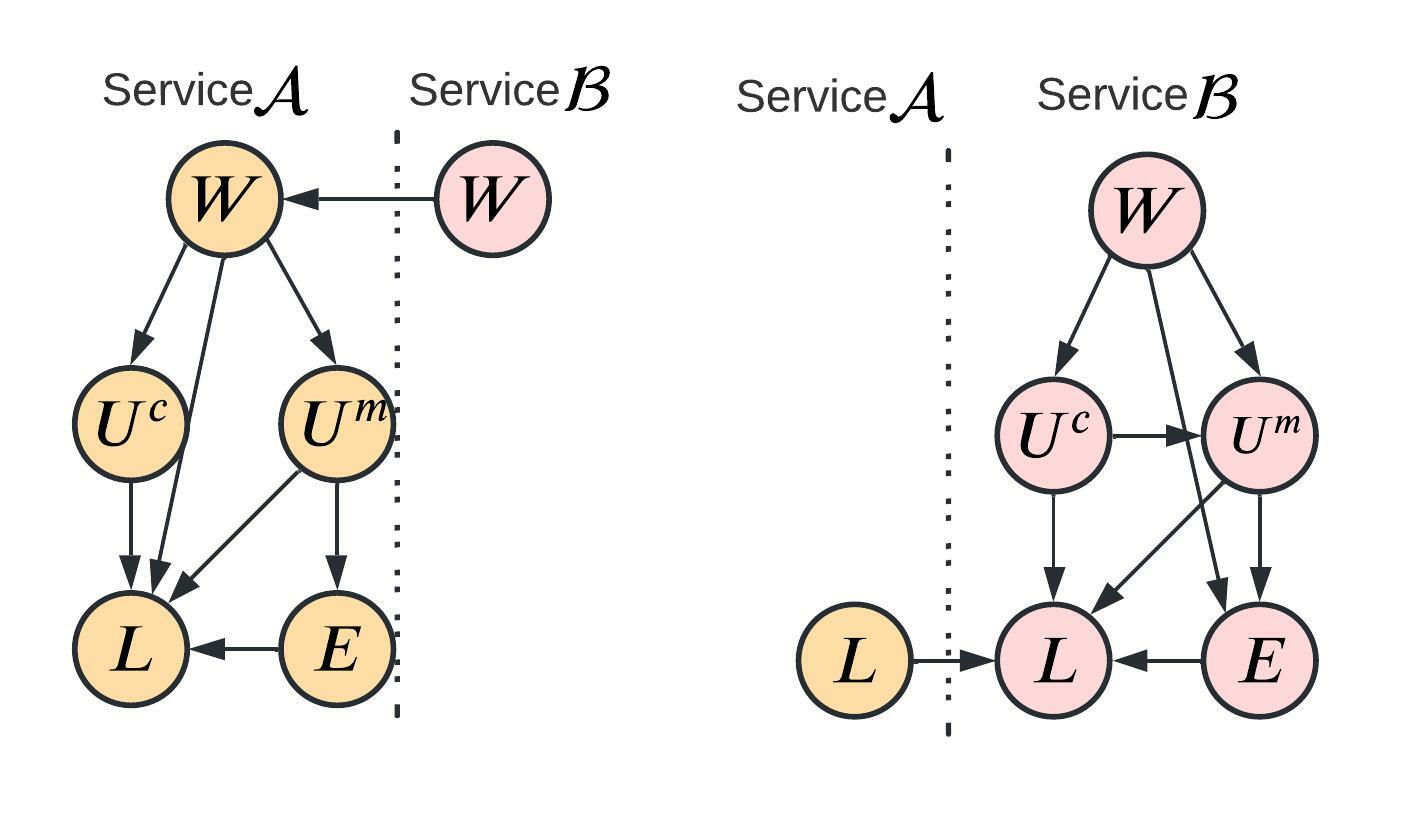}
         \caption{}
         \label{fig:individual-services-cg}
     \end{subfigure}
     \begin{subfigure}[b]{0.36\columnwidth}
         \centering
         \includegraphics[width=\columnwidth]{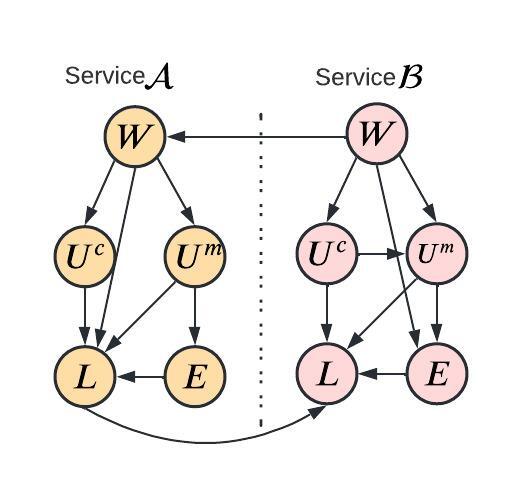}
         \caption{}
         \label{fig:complete-cg}
     \end{subfigure}
     \begin{subfigure}[b]{0.4\columnwidth}
         \centering
         \includegraphics[width=\columnwidth]{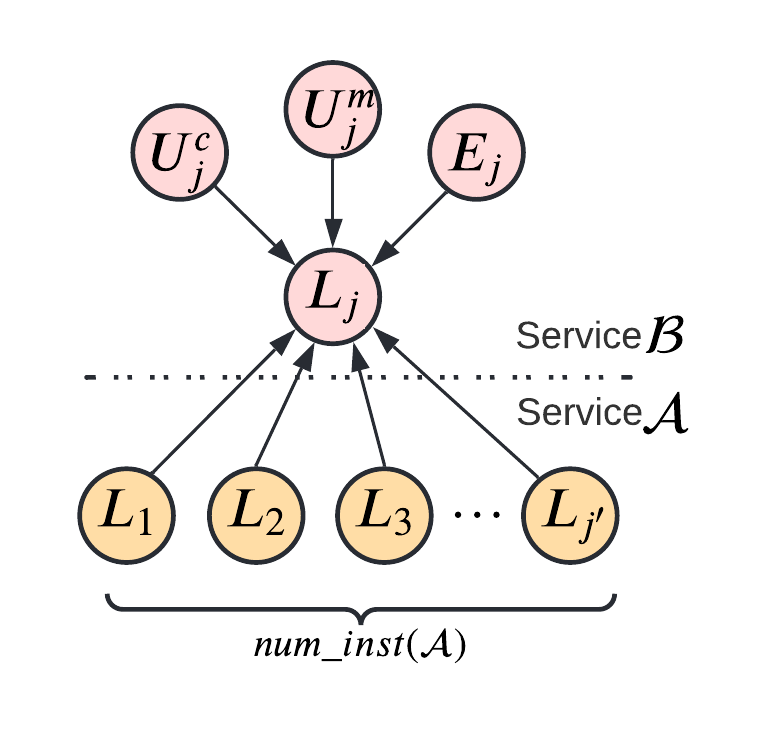}
         \caption{}
         \label{fig:multiple_instances_parents}
     \end{subfigure}
     \begin{subfigure}[b]{0.36\columnwidth}
         \centering
         \includegraphics[width=\columnwidth]{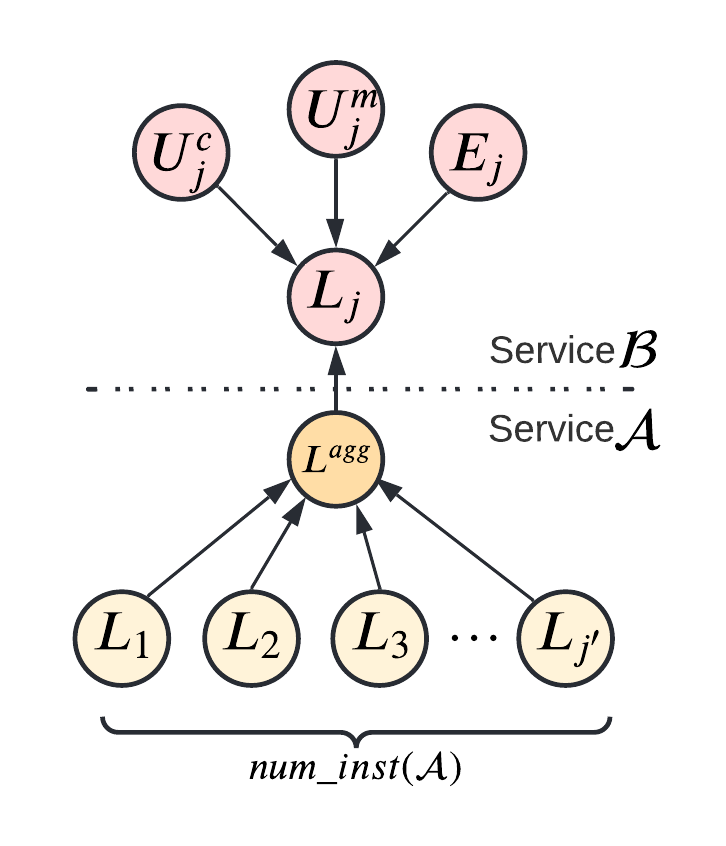}
         \caption{}
         \label{fig:aggrgegated_metric_parent}
     \end{subfigure}
    \Description[Examples of estimated causal graph for individual services A and B]{Figure (a) depicts an example of an estimated causal graph for individual services A and B, where service B calls service A. Figure (b) shows the merged causal graph, where error and workload are combined. Figures (c) and (d) represent the parent metrics for the latency of instance j of service A. In (c), the latency of service B is dependent on the metrics of service B and the latencies of all instances of service A. In (d), an aggregated latency node is constructed from the latencies of all instances of service A, serving as a latent node.}
    \caption{\textit{(a) shows an example estimated causal graph for individual services $\mathcal{A}$ and $\mathcal{B}$ where $\mathcal{B}$ calls $\mathcal{A}$. (b) shows the merged causal graph with error and workload merged. Figure (c) and (d) shows the parent metrics for latency of instance $j$ of $\mathcal{A}$. In (c) latency of $\mathcal{B}$ depends on metrics of $\mathcal{B}$ and latencies of all instances of $\mathcal{A}$. In (d) an aggregated latency node composed from the latencies of all instances of $\mathcal{A}$ is constructed, acting as a latent node.}}
    \label{fig:complete-individual-cg}
    \vspace{-0.6cm}
\end{figure}

\subsubsection{Structure Estimation}
Similar to Avg-fGES, \ourmethod{} identifies the causal structure between the metric nodes, as in \S \ref{sec:data_gen}, for one service at a time, making it highly scalable since its complexity will be linear in terms of the number of services. Causal structures of each microservice are then merged appropriately to form the final causal structure as shown in Fig. \ref{fig:complete-cg}. Following the conditional independence assumption of the instances at time $t$, metrics for each instance of a service can be treated as an iid sample from a distribution, realized by flattening the data over the instances.

To reduce computation and improve accuracy while estimating the set of parents for a metric, \ourmethod{} selects relevant parent metrics from the concerned service $\mathcal{S}$ and its adjacent services $\mathcal{S}_c$ that causally affects some metrics of $\mathcal{S}$ before running fGES. For any metric $x_{ijt}$ of instance $j$ of service $\mathcal{S}$, viable parent metrics set include all the metrics from the same instance, the workload for \textit{all instances} of caller services and latency and error metrics for \textit{all instances} of callee services of $\mathcal{S}$. For example, latencies of all instances of callee service can affect the latency of an instance of their caller (Fig. \ref{fig:multiple_instances_parents}). Metrics for all the instances belonging to only the adjacent services are aggregated for each service as illustrated in Fig. \ref{fig:aggrgegated_metric_parent}. This is supported by our domain knowledge, since all instances of caller service can call an instance of callee service, which gets managed by the load balancer. Thus, the possible metric nodes that can exhibit dependencies with instance $j$ of $\mathcal{S}$ are all the metrics $\{x_{ijt}\}, \forall i$, and union of the aggregation of all the metrics $i'$ over all instances of adjacent $\mathcal{S}_c$ that causally affect $\mathcal{S}$ in the interservice edges of the call graph, or $\{x^{agg}_{i't}\}$.





After filtering the data for each microservice $\mathcal{S}$, \ourmethod{} discovers the causal structure between the nodes. We minimize the BIC-based score function of fGES, which is computed by running a regression model with $x_{ijt}$ data flattened across number of instances as the dependent variable and parent metrics $\mathcal{P}(x_{ijt})$ as the predictors. The score function is defined as

\begin{equation} \label{eq:score}
\begin{split}
 Score(x_{ijt},\mathcal{P}(x_{ijt})) = & -2 \sum_{j,t} \log \mathcal(L (f_{i}(\mathcal{P}(x_{ijt})) | x_{ijt},\mathcal{P}(x_{ijt}))) \\
   & + \rho k \log n_i
\end{split}
\end{equation}
where, $\rho$ is a penalty term that penalizes the estimation of an over-fitting function with large number of parameters, $k$ number of parameters in $f_i$, and $n_i$ is the number of observation after flattening the data. We found $\rho = 2$ to be optimal for our experiments via hyperparameter tuning. 


\subsubsection{Structural Graph Post-Processing} \label{sec:cpdag}
\ourmethod{} generates a completed partially directed acyclic graph (CPDAG) comprising of directed as well as undirected edges. However, undirected edges render the causal graph infeasible to use for a downstream task like root cause analysis. Thus, we direct the undirected edges. To avoid cyclic dependency, we topologically sort the nodes and then assign the direction $n_i \rightarrow n_j$ where $i < j$ in topological ordering for the undirected edges. This prevents the cycles and results in a directed acyclic graph (DAG).

\subsubsection{Algorithm Specifics}
\vspace{-0.3cm}
\begin{algorithm}
\DontPrintSemicolon
\KwIn{Service Call Graph($G_{c}$), Metrics Data at instance level($\mathcal{D}$)}
\KwOut{Estimated Causal Graph}

Read Data $\mathcal{D}$\;
Read Service Call Graph $G_c$\;
$G_{out} \gets null$\;
\tcp*[l]{Iterate over each service}
\For{each service $N$}{
    $\mathcal{D}_{filtered} \gets $Data for metrics of all instances of service $N$\;
    $\mathcal{D}_{filtered} \gets $Agg `latency' and `error' over all instances for all child services of $N$\;
    $\mathcal{D}_{filtered} \gets $Agg `workload' over all instances for all parent services of $N$\;
    Stack data $\mathcal{D}_{filtered}$\;
    $G' \gets$ Run fGES on $\mathcal{D}_{filtered}$ with domain knowledge\;
    Append $G'$ to $G_{out}$\;
}
Connect inter-service edges in $G_{out}$\;
\Return{$G_{out}$}
\caption{Discover Causal Structure}

\label{alg:causil}
\end{algorithm}


Algorithm \ref{alg:causil} presents the workflow of \ourmethod{}. Note that we take only latency and error metrics for the child nodes and workload metric for the parent nodes due to the relationship stated in \S \ref{sec:problem-statement}. 

If the total latency of a request is larger than the data collection granularity, past workloads can affect metrics at the current timestamp. However, we observe that latency of requests is much lower (<100 ms) than the data collection granularity (15 min). Thus, past workloads on current system state is minimal and hence we do not consider the lags in metrics during structural graph construction.

\subsection{Incorporating Domain Knowledge} \label{sec:domain-knowledge}

Domain knowledge plays an important role in improving the performance of \ourmethod{} in terms of accuracy and computation time. System data and metric semantics provide several easy ways to generalize rules applicable to any microservice architecture without needing a system expert. We apply such rules to form a list of commonly prohibited edges. Nevertheless, any system expert can append to the domain knowledge already provided, which however can be time-consuming and hence we omit them in this paper. A prohibited edge list reduces the space-time complexity of the causal discovery algorithm and restricts the formation of unnecessary edges from system architecture point of view. We propose a list of rules based on Site Reliability Engineers' (SREs) input to automate the prohibited edge generation process.

Based on expert advice over the metric categories defined in \S \ref{sec:problem-statement}, we prohibit edges in the causal graph at a single service level such that: (i) No other metric within the same service for any instance can affect workload, it can be either an exogenous node or depend on the workload of caller service. (ii) latency cannot affect resource utilization. Furthermore, rules for prohibiting inter-service edges are: (i) Prohibit all edges between services if they are not connected in the call graph. (ii) Prohibit all edges between services that are connected except (a) Workload metric in the direction of the call graph and (b) Latency and Error metrics in the opposite direction of the call graph. This follows from the informed causal assumptions across microservices presented in \S \ref{sec:data_gen}. This list of prohibited edges for all the microservices serves as the domain knowledge to be introduced to the causal discovery algorithm.


\section{Experimental Setup} \label{sec:implement}

In this section, we give a brief description of the setup, dataset characteristics and the metrics used for the evaluation of our model. We implemented  \ourmethod{} in \textit{Python} while adapted and optimized publicly available libraries (\href{https://github.com/eberharf/fges-py }{fges-py} and \href{https://github.com/cmu-phil/tetrad}{tetrad}) for running fGES. We have run \ourmethod{} on a system having Intel Xeon E5-2686 v4 2.3GHz CPU with 8 cores.

\subsection{Baselines and Models}
We compare our proposed strategy \ourmethod{} against various baselines defined below. The algorithms below were tested \textit{with and without the application of Domain Knowledge (DK)}, which will be indicated as \textit{yes/no} in the evaluation tables.

\begin{enumerate}
\item \textbf{FCI: }Fast Causal Inference is a constraint-based algorithm algorithm~\cite{spirtes2000causation} which provides guarantees in causal discovery in the presence of confounding variables. We use the version of FCI that averages data across all instances.

\item \textbf{Avg-fGES\footnote{Comparison against baselines implementing other aggregation functions are reported in appendix \S \ref{sec:aggregation-baselines}}: }Implementation of the algorithm described in \S \ref{sec:averaging}. Multiple estimation functions $f_i$ in Eq. \ref{eq:condprobpool} estimating the dependency score between parent metrics and the child metric have been implemented; (i) \textbf{Avg-fGES-Lin}: Ordinary Least Square $f_i$, (ii) \textbf{Avg-fGES-Poly2}: Polynomial $f_i$ of degree 2, (iii) \textbf{Avg-fGES-Poly3}: Polynomial $f_i$ of degree 3.

\item \textbf{\ourmethod{}:} Different versions of the proposed method has been implemented that varies in their estimation functions $f_i$; 

(i) \textbf{\ourmethod{}-Lin}: Ordinary Least Square $f_i$, (ii) \textbf{\ourmethod{}-Poly2}: Polynomial $f_i$ of degree 2, (iii) \textbf{\ourmethod{}-Poly3}: Polynomial $f_i$ of degree 3.



\end{enumerate}

\subsection{Datasets} \label{sec:dataset}
We evaluated \ourmethod{} against a set of synthetic and semi-synthetic, while a case study on a real-world dataset is presented as well. 

\textit{Synthetic ($\mathcal{D}^{syn}$): }Three datasets are generated with 10/20/40 services and each service having 5 metric nodes, thus making a total of 50/100/200 nodes respectively in the causal graph. The metric nodes within and across the services were connected according to \S \ref{sec:data_gen}, and data was generated according to Alg. \ref{alg:distributed_data} in appendix, which accurately mimics the behavior of a load-balancer and auto-scaler in a real scenario. We present evaluations for each dataset by averaging across 5 graphs. For each instance of a service, time series dataset of a metric $\mathbf{x_{ijt}} = f_i(\mathcal{P}(\mathbf{x_{ijt}}))$ is generated, where $f$ is a non-linear quadratic function. Workload for a service $s$ ($W_s$) is distributed across each instance uniformly. Across adjacent services $a \rightarrow b$, $W_b = \beta W_a$, where $\beta \in [0,1]$.

\textit{Semi-Synthetic ($\mathcal{D}^{semi-syn}$): }With the same graphs generated as in above, causal relationships across metrics are learnt based on data gathered from a real-world service. For each edge in Figure \ref{fig:metric_graph}, we learn a random forest regressor and then generate output metric value with added random error given parent metric inputs to the function. The function learnt for a metric $x_{i}$ remains same for each instance and service. Workload for exogenous services is equal to the workload of real-world services.

\subsection{Evaluation Metrics}
Following existing works~\cite{raghu2018evaluation}, we evaluate the causal graph generated by \ourmethod{} using the following metrics.
\begin{enumerate}
    \item \textit{Adjacency Metrics (Adj)} denotes the correctness of the edges in the graph as estimated by \ourmethod{} by disregarding the directions of the edges.
    \item \textit{Arrow Head Metrics (AH)} considers the causal orientations of the edge when considered against the true graph and penalizes for the incorrectly identified directions of the correctly identified adjacencies.
    \item \textit{Structural Hamming Distance (SHD)} reports the number of changes that must be made to an estimated causal graph to recreate the ground truth graph.
\end{enumerate}
We report precision (P), recall (R) and F1-score (F) with their usual semantics for Adjacency and Arrow Head metrics. While adjacency metric records precision/recall of identified edges and rewards correctly identified edges, SHD measures edit distance between two graphs and penalizes misidentified/missing edges

\begin{figure}[t]
    \Description[Plots showing how domain knowledge helps in causal structure estimation]{Fig. (a), (c) and (e) are bar plots showing the difference in SHD for the three datasets (10/20/40 services). The difference between the run with domain knowledge shows a significant decrease in the SHD than the one run without any domain knowledge. Also, the time taken to construct the graph shows 70x improvement with domain knowledge.}
   \centering
     \begin{subfigure}[b]{0.49\columnwidth}
         \centering
         \includegraphics[width=0.9\columnwidth]{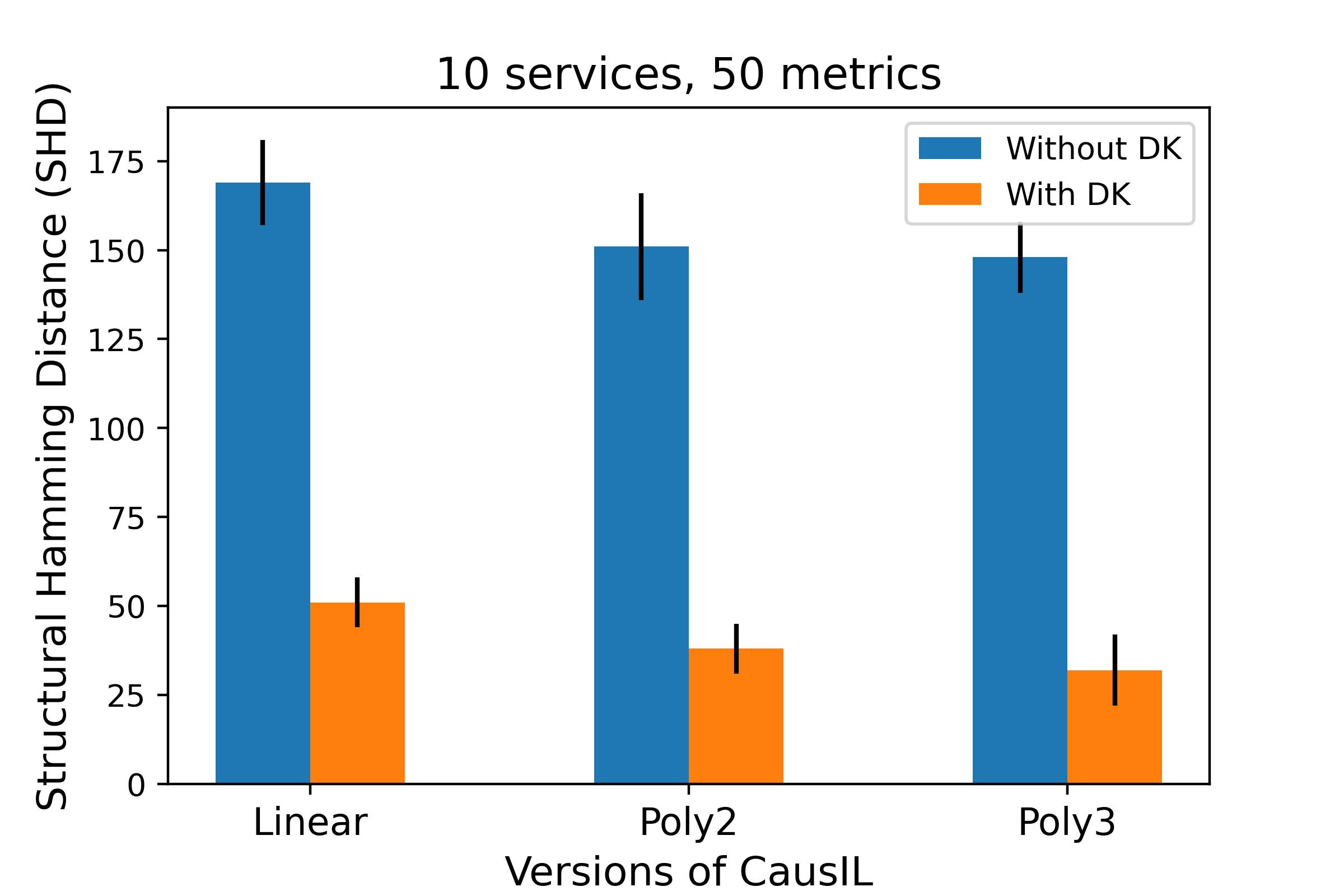}
         \caption{}
     \end{subfigure}
     \begin{subfigure}[b]{0.49\columnwidth}
         \centering
         \includegraphics[width=0.9\columnwidth]{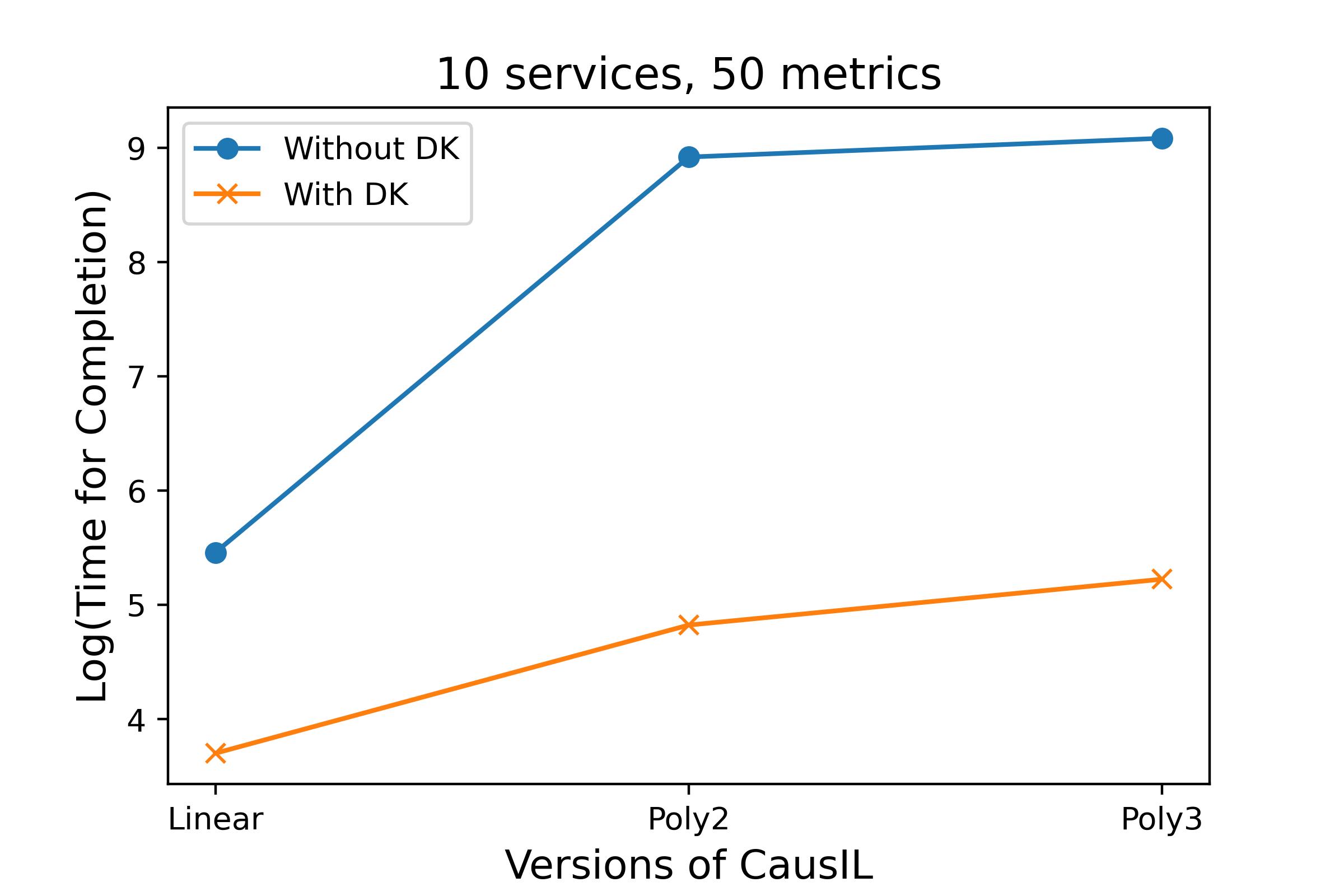}
         \caption{}
     \end{subfigure}
     \begin{subfigure}[b]{0.49\columnwidth}
         \centering
         \includegraphics[width=0.9\columnwidth]{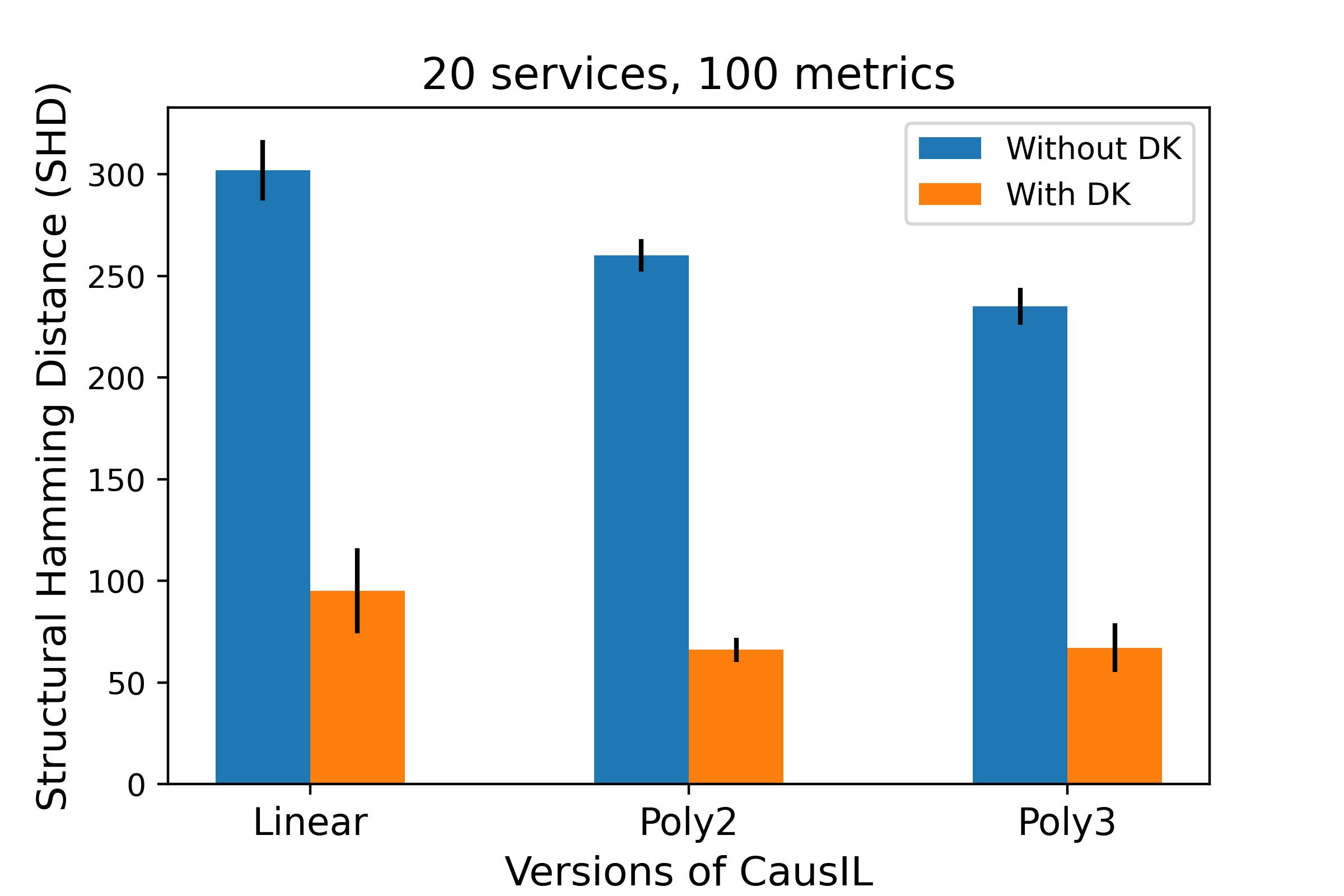}
         \caption{}
     \end{subfigure}
     \begin{subfigure}[b]{0.49\columnwidth}
         \centering
         \includegraphics[width=0.9\columnwidth]{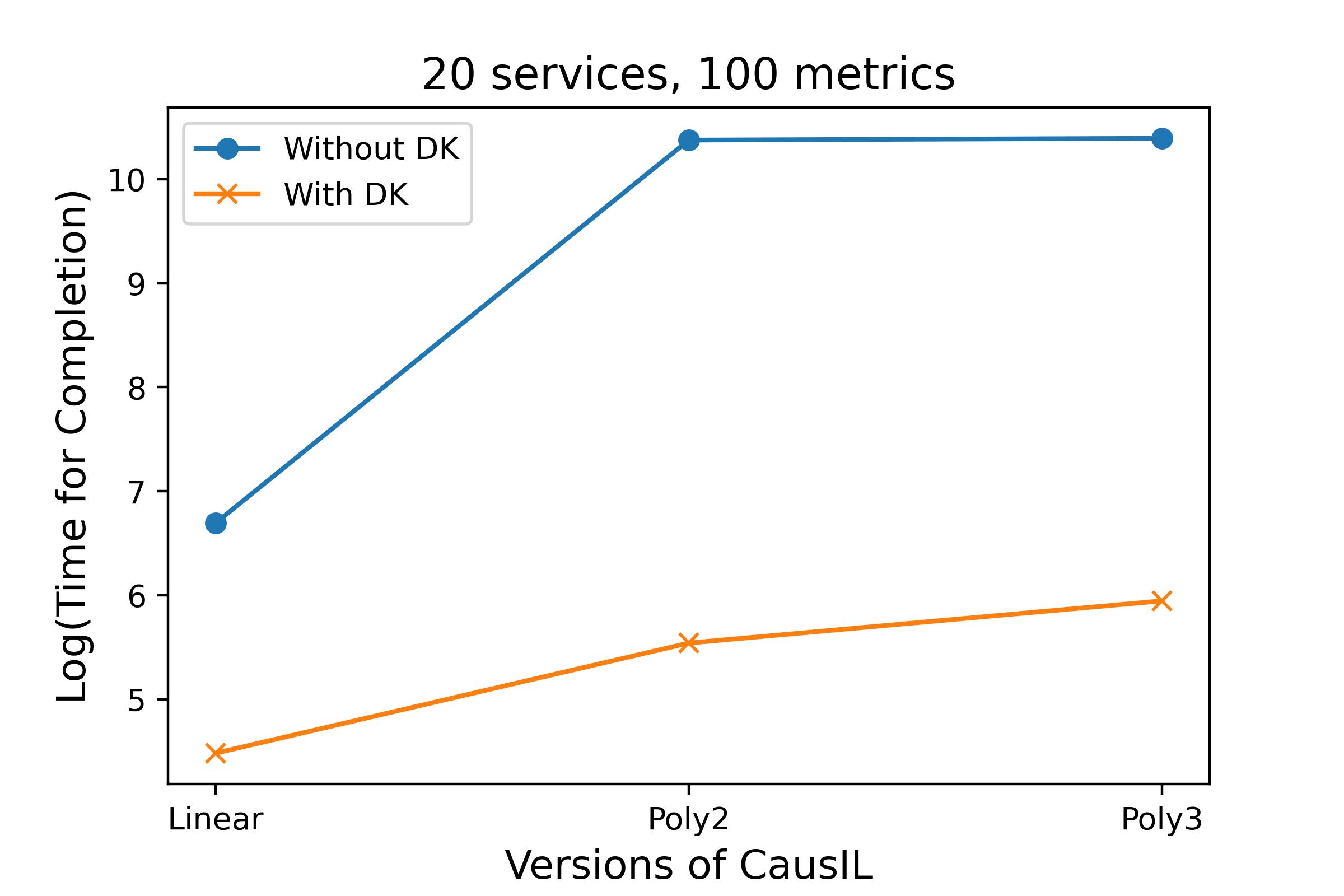}
         \caption{}
     \end{subfigure}
     \begin{subfigure}[b]{0.49\columnwidth}
         \centering
         \includegraphics[width=0.9\columnwidth]{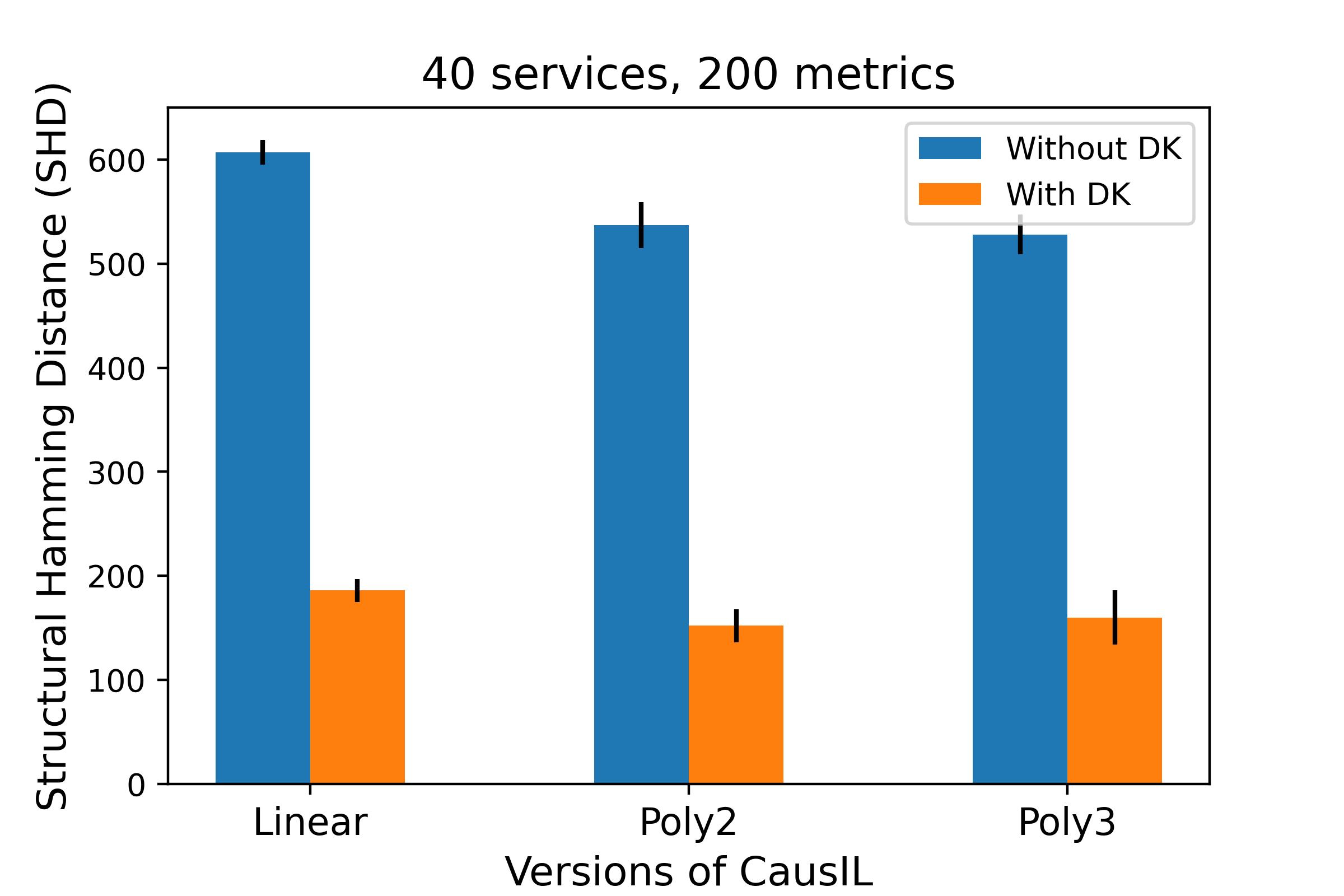}
         \caption{}
     \end{subfigure}
     \begin{subfigure}[b]{0.49\columnwidth}
         \centering
         \includegraphics[width=0.9\columnwidth]{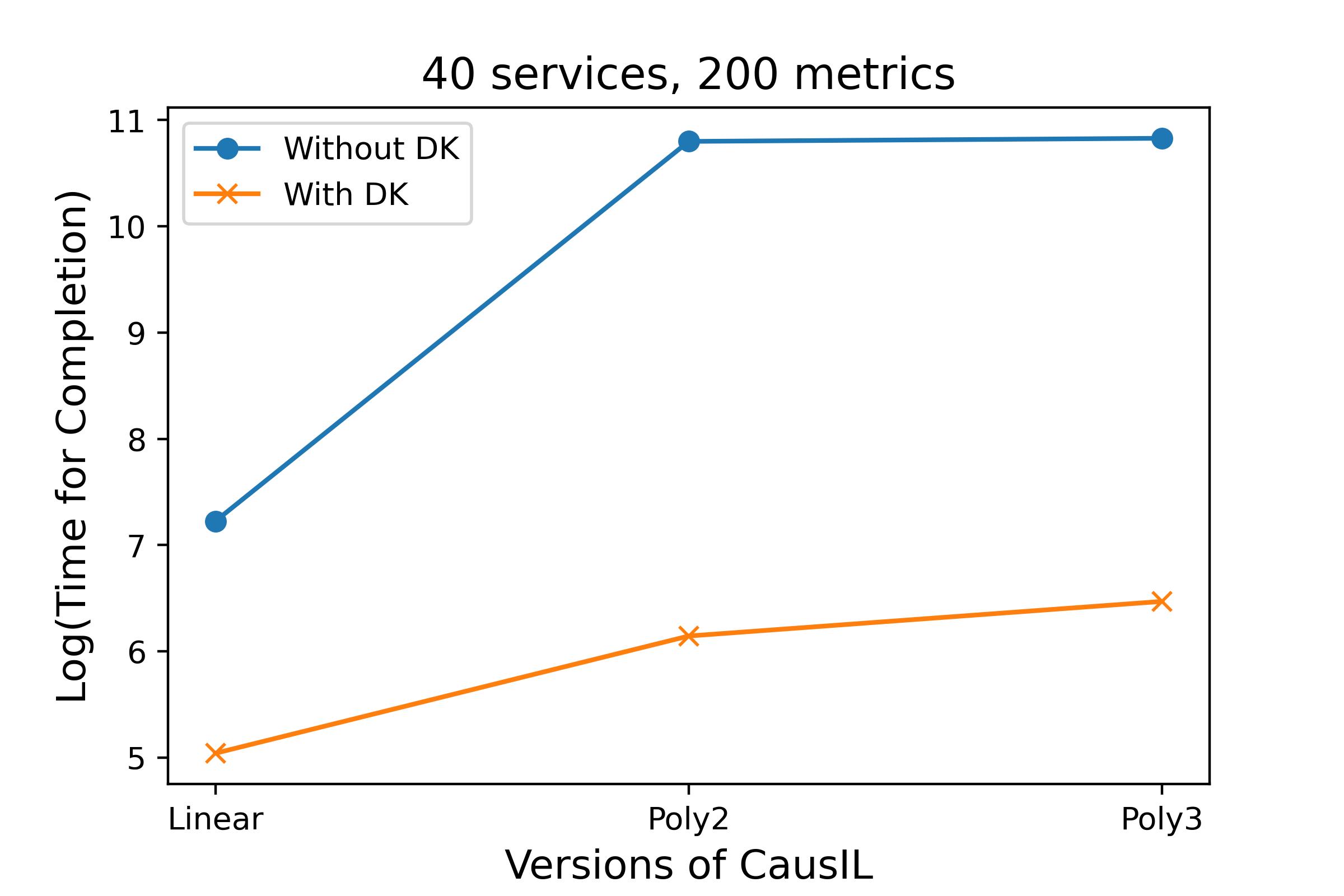}
         \caption{}
     \end{subfigure}
    \vspace{-0.1 cm}
    \caption{\textit{Demonstrating how domain knowledge (DK) helps in causal structure estimation. Fig. (a), (c) and (e) show the difference in SHD for the three datasets, while (b), (d) and (f) show the improvements in time (logarithmic scale).}}
    \vspace{-0.4 cm}
    
    \label{fig:dk}
    
\end{figure}

\section{Evaluation Results} \label{sec:results}

This section presents the evaluations performed to demonstrate the effectiveness of \ourmethod{} against the various baselines.
\vspace{-0.1cm}
\subsection{Impact of Domain Knowledge}


We state in \S \ref{sec:domain-knowledge} that the use of domain knowledge for the system architecture in the form of prohibited edge list greatly boosts the accuracy of \ourmethod{}. Additionally, it also reduces computational time and complexity. In this section, we elucidate this improvement in performance with synthetic data for different version of \ourmethod{} in Fig. \ref{fig:dk}. We only report SHD in this section since it can effectively penalize redundant edges estimated when \ourmethod{} is employed without domain knowledge. Thus it forms a more decisive metric than the others in evaluating the impact of domain knowledge in the construction of the causal graph.


Fig. \ref{fig:dk} shows that using domain knowledge of the system architectures improves the SHD of the estimated causal graph by more than $3.5\times$ across all models. With domain knowledge restricting the formation of edges that violate system rules, \ourmethod{} can better understand the structure of the system metrics and keep only the relevant edges. It also reduces the time taken for \ourmethod{} to estimate the causal graph by more than $70\times$ because computing the causal dependency score and function estimation by the underlying fGES is not required for certain child-parent pairs. Also, because the dataset was generated using a non-linear function, the model version implementing the linear estimation function performs the worst for each dataset. The benefit of introducing domain knowledge can be observed across various scales of the dataset. We will report the results of the remaining evaluations when estimation involved domain knowledge.

\subsection{Baseline Comparison} \label{sec:baseline}

\begin{table*}

\centering
\resizebox{0.85\textwidth}{!}{
\begin{tabular}{c|c|ccccccc|ccccccc}
\toprule
\multirow{2}{*}{\begin{tabular}[c]{@{}c@{}} \# Services, \\ \# Metrics \end{tabular}} & \multirow{2}{*}{\begin{tabular}[c]{@{}c@{}} Model \end{tabular}} & \multicolumn{7}{c|}{$\mathcal{D}^{syn}$} & \multicolumn{7}{c}{$\mathcal{D}^{semi-syn}$} \\ 
\cmidrule{3-16}
  &   & SHD & AdjP & AdjR & AdjF & AHP & AHR & AHF & SHD  & AdjP & AdjR & AdjF & AHP & AHR & AHF \\

\midrule
\midrule
\multirow{7}{*}{\begin{tabular}[c]{@{}c@{}} 10, 50 \end{tabular}}  & FCI &  53 & 0.772  &  0.841  &  0.805  &  0.704  &  \textbf{0.909}  &  0.793 & 53 & 0.762  &  0.873  &  0.814  &  0.69  &  \textbf{0.906}  &  0.783  \\
\cmidrule{2-16}
  & Avg-fGES-Lin &  54 & 0.794  &  0.854  &  0.822  &  0.686  &  0.864  &  0.765 & 54 & 0.793  &  0.851  &  0.82  &  0.68  &  0.857  &  0.759 \\
\cmidrule{2-16}
& Avg-fGES-Poly2 &  48 & 0.81  &  0.861  &  0.834  &  0.723  &  0.892  &  0.799 & 50 & 0.807  &  0.845  &  0.825  &  0.713  &  0.883  &  0.789 \\
\cmidrule{2-16}
& Avg-fGES-Poly3 &  46 & 0.837  &  0.839  &  0.837  &  0.747  &  0.893 &  0.814 & 46 & 0.837  &  0.838  &  0.836  &  0.745  &  0.89 &  0.811 \\
\cmidrule{2-16}
& \ourmethod{}-Lin &  51 & 0.788  &  0.878  &  0.83  &  0.695  &  0.882  &  0.777 & 53 & 0.788  &  0.874  &  0.829  &  0.684  &  0.868  &  0.765 \\
\cmidrule{2-16}
& \ourmethod{}-Poly2 &  38 & 0.889  &  0.852  &  0.869  &  0.795  &  0.895  &  0.842 & 36 & 0.892  &  \textbf{0.877}  &  0.883  &  0.795  & 0.892  &  \textbf{0.84} \\
\cmidrule{2-16}
& \ourmethod{}-Poly3 &  \textbf{32} & \textbf{0.909}  &  \textbf{0.878}  &  \textbf{0.891}  &  \textbf{0.823}  &  0.905  &  \textbf{0.862} & \textbf{35} & \textbf{0.909}  &  0.875  &  \textbf{0.89} &  \textbf{0.797}  &  0.877  &  0.835 \\
\midrule
\midrule
\multirow{7}{*}{\begin{tabular}[c]{@{}c@{}} 20, 100 \end{tabular}}  & FCI &  105  &  0.773  &  0.85  &  0.81  &  0.702  &  \textbf{0.909}  &  0.792 & 105 & 0.772 & 0.856 & 0.811 & 0.699 & \textbf{0.906} & 0.79
 \\
\cmidrule{2-16}
  & Avg-fGES-Lin &  103 & 0.814  &  0.832  &  0.82  &  0.706  &  0.867  &  0.778 & 104 & 0.813  &  0.827  &  0.818  &  0.709  &  0.872  &  0.782 \\
\cmidrule{2-16}
& Avg-fGES-Poly2 &  95 & 0.823  &  0.872  &  0.847  &  0.716  &  0.87  &  0.786 & 93 & 0.826  &  0.886  &  0.855  &  0.713  &  0.864  &  0.781 \\
\cmidrule{2-16}
& Avg-fGES-Poly3 &  88 & 0.845  &  0.867  &  0.857  &  0.74  &  0.876  &  0.802 & 85 & 0.846  &  0.871  &  0.836  &  0.782  &  0.869  &  0.798 \\
\cmidrule{2-16}
& \ourmethod{}-Lin &  95 & 0.812  &  0.856  &  0.832  &  0.728  &  0.897  &  0.803 & 100 & 0.809  &  0.836  &  0.82  &  0.722  &  0.892  &  0.797 \\
\cmidrule{2-16}
& \ourmethod{}-Poly2 &  \textbf{66} & 0.908  &  \textbf{0.895}  &  \textbf{0.901}  &  0.801  &  0.883  &  0.84 & \textbf{68} & 0.907  &  \textbf{0.892}  &  \textbf{0.899}  &  0.795  &  0.876  &  0.833 \\
\cmidrule{2-16}
& \ourmethod{}-Poly3 &  67 & \textbf{0.913}  &  0.881  &  0.896  &  \textbf{0.807}  &  0.884  &  \textbf{0.844} & 72 & \textbf{0.911}  &  0.859  &  0.884  & \textbf{ 0.804}  &  0.882  &  \textbf{0.841} \\
\midrule
\midrule
\multirow{7}{*}{\begin{tabular}[c]{@{}c@{}} 40, 200 \end{tabular}}  & FCI &  204 & 0.792 & 0.814 & 0.803 & 0.719 & 0.907 & 0.803 & 206 & 0.781 & 0.802 & 0.791 & 0.706 & \textbf{0.904} & 0.793 \\
\cmidrule{2-16}
  & Avg-fGES-Lin &  203 & 0.837  &  0.78  &  0.807  &  0.742  &  0.886  &  0.808 & 206 & 0.837  &  0.778  &  0.806  &  0.737  &  0.88  &  0.802 \\
\cmidrule{2-16}
& Avg-fGES-Poly2 &  197 & 0.853  &  0.782  &  0.815  &  0.749  &  0.879  &  0.809 & 200 & 0.852  &  0.779  &  0.813  &  0.746  &  0.876  &  0.806 \\
\cmidrule{2-16}
& Avg-fGES-Poly3 &  204 & 0.862  &  0.741  &  0.795  &  0.763  &  0.886  &  0.82 & 203 & 0.862  &  0.746  &  0.799  &  0.759  &  0.882  &  0.816 \\
\cmidrule{2-16}
& \ourmethod{}-Lin &  186 & 0.835  &  0.811  &  0.824  &  0.759  &  0.897  &  0.827 & 188 & 0.833  &  0.811  &  0.822  &  0.755  &  0.905 &  0.823 \\
\cmidrule{2-16}
& \ourmethod{}-Poly2 &  \textbf{152} & 0.919  &  \textbf{0.828}  &  \textbf{0.871}  &  0.825  &  0.899  &  0.86 & \textbf{155} & 0.92  &  \textbf{0.817}  &  \textbf{0.864}  &  0.809  &  0.879  &  0.843 \\
\cmidrule{2-16}
& \ourmethod{}-Poly3 &  160 & \textbf{0.922}  &  0.783  &  0.846  &  \textbf{0.83}  &  \textbf{0.909}  &  \textbf{0.863} & 162 & \textbf{0.922}  &  0.784  &  0.847  &  \textbf{0.821}  &  0.89  &  \textbf{0.854} \\
\bottomrule
\end{tabular}
}
\Description[Table showing the comparison of \ourmethod{} against the baselines for all types of datasets.]{
It summarizes the performance of CausIL for synthetic and
semi-synthetic data and demonstrates how preserving the instance-specific variations across various services helps and all estimation functions of CausIL outperform the baselines.
}
\caption{Comparison of \ourmethod{} against the baselines for all types of datasets.}
\vspace{-0.4cm}
\label{tab:baseline}
\end{table*}

Table \ref{tab:baseline} summarizes the performance of \ourmethod{} for synthetic and semi-synthetic data and demonstrates how preserving the instance-specific variations across various services helps. The estimated causal graph is evaluated against the ground truth causal graph constructed during data generation, and thus, \ourmethod{} aims to implicitly learn the data generation process.

We observe from Table \ref{tab:baseline} that all estimation functions of \ourmethod{} outperform their corresponding versions of Avg-fGES by $\sim25\%$ on average. \ourmethod{} captures more granular information about the system structure and how instance-specific variation of each metric causally affects other metrics. Thus, the causal graph learned by \ourmethod{} is more accurate than the baselines when compared against ground-truth (more correctly identified edges and fewer missing edges). We demonstrate this using a toy example (\S \ref{sec:toy}), where Avg-fGES could not discover (cpu utilization, latency) and (memory, error) edges, which \ourmethod{} predicted. Qualitatively, for downstream tasks like root cause detection, inaccurate graph estimated by the baselines can lead to incorrect conclusions about faulty service if graph traversal algorithms~\cite{chen2014causeinfer} are used, which a graph estimated by \ourmethod{} can limit.

Furthermore, for each of the baseline categories, the one implementing the non-linear estimation function provides better metric values than the linear counterpart. With data being generated from a non-linear function, it is intuitive that a linear estimation function won't be able to capture the relationships succinctly. With increase in the degree of non-linearity in estimation function, \ourmethod{} shows a general trend in performance. We have implemented multiple versions of polynomial estimation functions $f_i$ differing in their degree. However, a score-based approach (BIC) can also guide the function choice. \ourmethod{} also outperforms FCI in SHD. We notice that arrow head recall for FCI is more than \ourmethod{}, though precision is less. This implies that FCI generates a causal graph with a large number of directed edges as compared to ground truth, trying to estimate a denser graph. In terms of memory overhead, \ourmethod{} incurs ~100MB more than baseline on average.

We further observe that \ourmethod{} with synthetic data outperforms when run on semi-synthetic data for the same graph and the same model. This is because semi-synthetic data have correlated and dependent values among various exogenous nodes as well as maintain autoregressive nature of the time series, which makes it intrinsically harder for an algorithm to establish causal dependencies.


\vspace{-0.3cm}
\subsection{Real-World Use Case Study}
We further evaluate \ourmethod{} against a real-world dataset collected from a part of a production-based microservice system of an enterprise cloud-service. The call graph for the system follows a star-shaped architecture with a total of 1 monolith calling 9 microservices (\S \ref{sec:real-graph}). The data was collected from Grafana, a monitoring tool that tracks and logs the metric values of several components of a running system at certain time intervals. We collected data for a span of 2 months with 5 minutes granularity.

With the unavailability of a ground truth causal graph between performance metrics for real data, we construct it based on the causal assumptions stated in \S \ref{sec:data_gen} and then evaluate the estimated causal graph against it. We observe from Table \ref{tab:RealDataTable} that \ourmethod{}-Poly2 performs the best among the other models. Though versions of Avg-fGES have high adjacency values, that is, they can estimate edges between the metrics, but the direction of the edges suffers, which is evident from their low arrow head metrics. FCI also performs poorly with real data. Each of the models on average estimates 25-30 undirected edges out of possible 113 ground truth edges.

\begin{table}[h]
\centering
\resizebox{0.9\columnwidth}{!}{
\begin{tabular}{c|ccccccc}
\toprule
Model & SHD & AdjP & AdjR & AdjF & AHP & AHR & AHF \\
\midrule
\midrule
FCI & 59  &  0.756  &  0.796  &  0.775  &  0.697  &  0.922  &  0.794  \\
\midrule
Avg-fGES-Lin & 52 & 0.829 & 0.858 & 0.843 & 0.692 & 0.835 & 0.757  \\
\midrule
Avg-fGES-Poly2 & 53 & 0.823 & 0.823 & 0.823 & 0.708 & 0.86 & 0.777  \\
\midrule
Avg-fGES-Poly3 & 51 & \textbf{0.852} & 0.814 & 0.833 & 0.722 & 0.848 & 0.78  \\
\midrule
\ourmethod{}-Lin & 50 & 0.807 & 0.814 & 0.81 & 0.737 & 0.913 & 0.816  \\
\midrule
\ourmethod{}-Poly2 & \textbf{40} & 0.818 & \textbf{0.876} & \textbf{0.846} & \textbf{0.785} & \textbf{0.96} & \textbf{0.864}  \\
\midrule
\ourmethod{}-Poly3 & 46 & 0.824 & 0.867 & 0.845 & 0.739 & 0.898 & 0.811  \\
\bottomrule
\end{tabular}
}
\Description[Table showing the results of experiments on real data]{Comparison shows that CausIL-Poly2 performs the best among the other models when run on real data}
\caption{Experiments on Real Data}
\vspace{-0.5cm}
\label{tab:RealDataTable}
\end{table}

\vspace{-0.2cm}
\section{Discussion} \label{sec:discussion}


\textbf{Scalability:} \ourmethod{} discovers service-specific causal structures and then combines them to form the entire causal graph. Adding a new microservice will scale the computation time linearly. Based on the experiments using \ourmethod{}-Poly2 on multiple datasets, the average time for causal discovery for each service is 12-13s, with a standard deviation of 1.09s. Furthermore, parallelizing the discovery for each microservice would further improve the computation time.

\textbf{Domain Knowledge:}  Aggregating minimal domain knowledge is not expensive since the edge exclusion list that \ourmethod{} uses is based on generic system architecture rules and principles. However, system engineers can decide to modify the rules as well, which might be expensive but not necessary, hence not compromising the estimation accuracy. 
Though argued to be a bad architectural design~\cite{taibi2020microservices}, some real systems might also exhibit cyclic dependencies \cite{luo2021characterizing}, making it hard to model. However, even in the presence of circular requests, microservice operations differ either in terms of multiple regional placements or distinct components are invoked. In such cases, we can essentially split it up into sub-services performing individual operations which can easily be handled by \ourmethod{}. On the contrary, circular dependency on services having identical operations will lead to an infinite loop and hence is not observed in real systems.

\vspace{-0.2cm}
\section{Conclusion}
In this paper, we present a novel causal structure detection methodology \ourmethod{} that leverages metric variations from all the instances deployed per microservice. It makes a practical assumption based on system domain knowledge that the multiple instances of a service are identical and independent to each other conditioned on the load request received. Thus, \ourmethod{} filters relevant system metrics and models the causal graph at the metric-level using instance-level data variations. It estimates a causal graph for each microservice individually and then aggregates them over all microservices to form the final graph. An added advantage is its capability to cope with instances’ distinct and transitional nature, usually observed in a microservice deployment due to auto-scaler configuration.

We further show that incorporating system domain knowledge improves causal structure detection in terms of accuracy and computation time. It helps in estimating the essential edges and ignoring the non-existential edges. From our evaluations on simulated data, we show that our method outperforms the baselines by $\sim25\%$, and the introduction of domain knowledge improves SHD by $\sim3.5\times$ on average. We also evaluate on real data elucidating the practical usefulness of \ourmethod{}. 



\bibliographystyle{ACM-Reference-Format}
\bibliography{main}

\appendix

\begin{figure*}[t]
    \centering
    \includegraphics[width=0.8\textwidth]{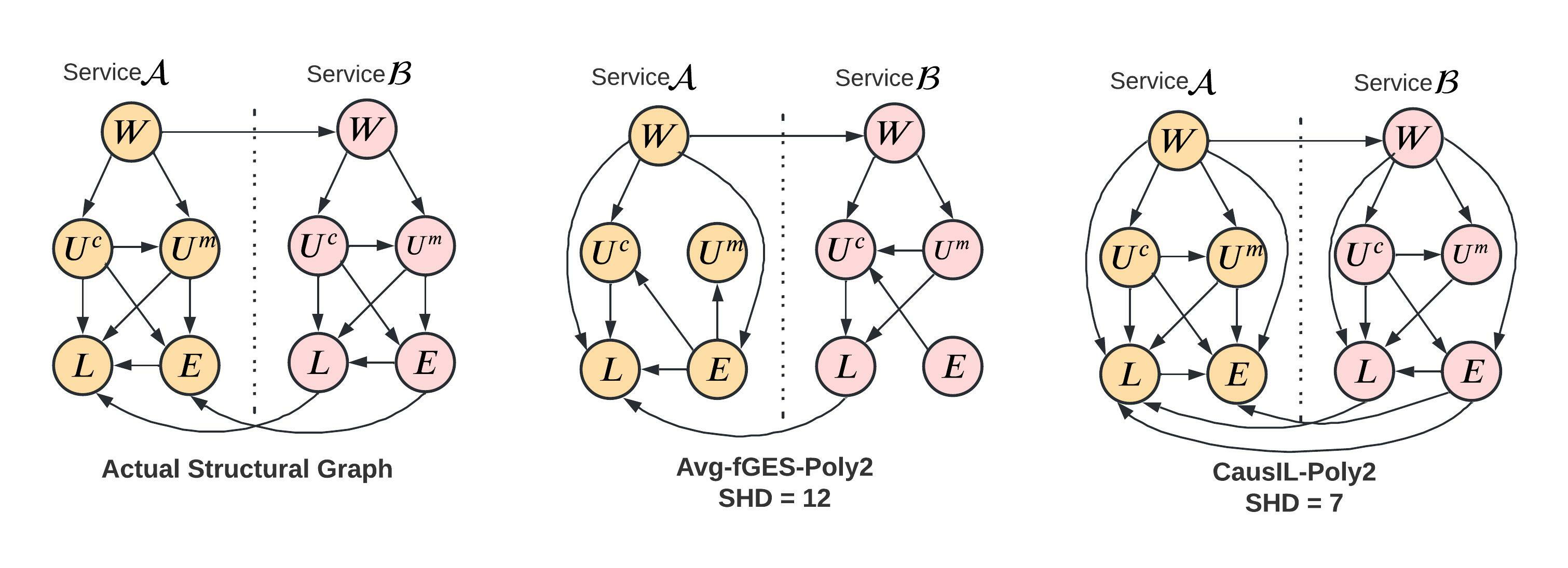}
    \vspace{-0.8cm}
    \Description[Example on a dummy setup to show the benefit of CausIL]{Example shows that the estimated causal graph from CausIL is able to identify more edges correctly as compared to the Avg-fGES on a 2 service dummy setup.}
    \caption{Illustration of the benefits of \ourmethod{} on a toy graph.}
    \label{fig:toyexample}
\end{figure*}

\section{Additional Evaluations}
\subsection{Illustration on Toy Graph} \label{sec:toy}
We illustrate why \ourmethod{} works better than the baseline via an example of a small toy call graph (Figure \ref{fig:toyexample}). We consider two services $A$ and $B$, such that $A$ calls $B$ ($A \rightarrow B$). For each of the service, we generate synthetic data (\S \ref{sec:dataset}) for the metrics categories listed in \S \ref{sec:data_gen} and create a ground truth causal graph following the causal assumptions. We run \ourmethod{}-Poly2 and Avg-fGES-Poly2 with domain knowledge on the synthetically generated data and report our findings here.

We observe that the Structural Hamming Distance for the estimated causal graph with \ourmethod{} is lower than the one estimated with the baseline method. Furthermore, some edges were not identified by Avg-fGES, which was identified by \ourmethod{}. This is because of the non-linearity in the dataset is reduced when the corresponding metrics for multiple instances are averaged. It does not capture the true distirbution of the data generation process. An evident example for this is the edge from memory utilization to latency for service $A$, which has been identified by \ourmethod{}, but Avg-fGES misses it. The adjacency and arrow head F1 score for \ourmethod{} are 0.857 and 0.774 respectively, while the same for Avg-fGES are 0.645 and 0.636 respectively, furthering bolstering the efficacy of \ourmethod{}.

\subsection{Comparison against multiple aggregation function-based baselines} \label{sec:aggregation-baselines}
In \S \ref{sec:baseline}, we have evaluated and compared \ourmethod{} against Avg-fGES, which aggregates a particular metric values for all the instances running for a service at any time $t$ by averaging them. However, though we argue that any aggregation function like summation, maximum etc. will have the same shortcoming as doing averaging where the entire spectrum of the relationship will not be captured, we empirically show this in Table \ref{tab:all_func}.

We have evaluated on the synthetic dataset generated at multiple scales, that is, with 50 metric nodes, 100 metric nodes and 200 metric nodes. We report SHD, adjacency F1 score and Arrow Head F1 score. The baselines that we have evaluated against are:
\begin{enumerate}
    \item Avg-fGES, where metric values over all instances are averaged at any particular time $t$
    \item Max-fGES, where the maximum metric value over all the instances is taken at any time instant $t$
    \item Min-fGES, where the minimum metric value is chosen
    \item Sum-fGES, where metric values for all the instances are summed at any time instant $t$
\end{enumerate}

We observe that \ourmethod{} outperforms all the baselines with Avg-fGES and Sum-fGES being the better performing baselines. This proves our claim empirically and the need for the design of a causal structure estimation method that takes into account instance specific variations.

\begin{table*}
\centering
\resizebox{\textwidth}{!}{
\begin{tabular}{c|c|ccc|ccc|ccc|ccc|ccc}
\toprule
\multirow{2}{*}{\begin{tabular}[c]{@{}c@{}} \# Services, \\ \# Metrics \end{tabular}} & \multirow{2}{*}{\begin{tabular}[c]{@{}c@{}} Estimation\\ Function \end{tabular}} & \multicolumn{3}{c|}{Avg-fGES} & \multicolumn{3}{c|}{Max-fGES} & \multicolumn{3}{c|}{Min-fGES} & \multicolumn{3}{c|}{Sum-fGES} & \multicolumn{3}{c}{\ourmethod{}} \\ 
\cmidrule{3-17}
  &   & SHD & AdjF & AHF & SHD & AdjF & AHF & SHD & AdjF & AHF & SHD & AdjF & AHF & SHD & AdjF & AHF \\
\midrule
\midrule
\multirow{3}{*}{\begin{tabular}[c]{@{}c@{}} 10, 50 \end{tabular}}  & Linear &  54  &  0.822  &  0.765  &  53  &  0.832 &  0.749 &  56  &  0.796  &  0.779  &  53  &  0.804  &  0.79  &  51  &  0.83  &  0.777   \\
\cmidrule{2-17}
  & Poly2 &  48  &  0.834  &  0.799  &  53  &  0.824  &  0.765  &  53  &  0.813  &  0.778  &  50  &  0.822  &  0.8  &  38  &  0.869  &  0.842   \\
\cmidrule{2-17}
  & Poly3 &  46  &  0.837  &  0.814  &  56  &  0.797  &  0.776  &  50  &  0.801  &  0.817  &  47  &  0.825  &  0.821  &  32  &  0.891  &  0.862   \\
\midrule
\midrule
\multirow{3}{*}{\begin{tabular}[c]{@{}c@{}} 20, 100 \end{tabular}}  & Linear &  103  &  0.82  &  0.778  &  98  &  0.834  &  0.785  &  109  &  0.807  &  0.775 &  101  &  0.816  &  0.801  &  95  &  0.832  &  0.803   \\
\cmidrule{2-17}
  & Poly2 &  95  &  0.847  &  0.786  &  97  &  0.827  &  0.802  &  99  &  0.822  &  0.801  &  89  &  0.853  &  0.799  &  66  &  0.901  &  0.84   \\
\cmidrule{2-17}
  & Poly3 &  88  &  0.857  &  0.802  &  106  &  0.813  &  0.778  &  106  &  0.797  &  0.801  &  86  &  0.847  &  0.825  &  67  &  0.896  &  0.844   \\
\midrule
\midrule
\multirow{3}{*}{\begin{tabular}[c]{@{}c@{}} 40, 200 \end{tabular}}  & Linear &  203  &  0.807  &  0.808  &  204  &  0.802  &  0.812  &  229  &  0.765  &  0.802  &  211  &  0.791  &  0.812  &  186  &  0.824  &  0.827   \\
\cmidrule{2-17}
  & Poly2 &  197  &  0.815  &  0.809  &  211  &  0.789  &  0.813  &  228  &  0.76  &  0.809  &  205  &  0.807  &  0.802  &  152  &  0.861  &  0.86   \\
\cmidrule{2-17}
  & Poly3 &  204  &  0.795  &  0.82  &  214  &  0.777  &  0.821  &  224  &  0.757  &  0.823  &  204  &  0.803  &  0.808  &  160  &  0.846  &  0.863   \\

\bottomrule
\end{tabular}
}
\Description[Table showing the results of experiments based on different aggregation function]{Comparison shows that CausIL outperforms all the versions of instance level data aggregation schemes like Avg-fGES, Sum-fGES, Max-fGES and Min-fGES.}
\caption{Comparison of \ourmethod{} against multiple baselines. Avg-fGES averages the metric values over all instances for each time instant. Similarly, Max-fGES computes the maximum, Min-fGES computes the minimum and Sum-fGES computes the sum of the metric values over all instances for any time $t$. }
\label{tab:all_func}
\end{table*}

\section{Real Graph Details} \label{sec:real-graph}

\begin{figure}[H]
    \centering
    \includegraphics[width=0.4\textwidth]{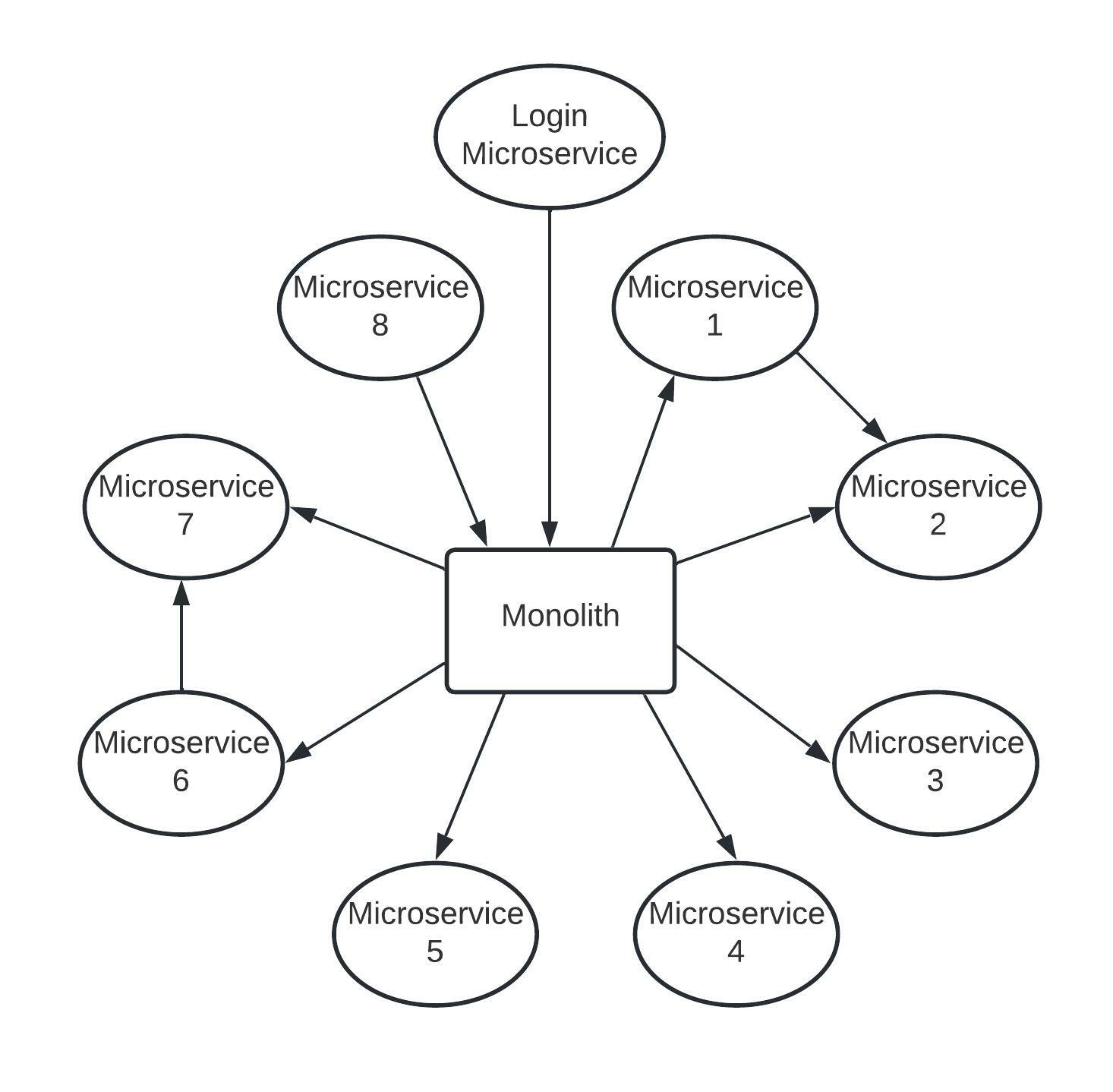}
    \Description[Service Call Graph for real data]{Total of 9 microservices in a star schema where the monolith calls 8 microservices and 1 microservice calls the monolith. Apart from these, microservice 1 also calls microservice 2, and microservice 6 calls microservice 7.}
    \caption{Service Call Graph for real data}
    \label{fig:realGraph}
\end{figure}

In our study with the real data, the underlying service architecture from which the data was collected is illustrated in Figure \ref{fig:realGraph}. We collect metrics for each service belonging to the metric categories defined in \S \ref{sec:data_gen}, and construct a causal structural graph against which we evaluate. Though, ground truth graph is unknown in a real world setting, we use our constructed graph as a proxy for the ground truth.

\newpage
\section{Data Generation}
From a given structural graph, we generate synthetic and semi-synthetic data. We construct a directed acyclic graph of services (call graph) and use it to generate a ground truth causal graph at the metrics level, following existing work~\cite{zheng2018dags}. The data is generated by using the edges between performance metrics for each service. Algorithm \ref{alg:generate_graph} is used to generate a random graph between services, which is then trivially extended to form the graph between performance metrics using the causal assumptions and rules described in \S \ref{sec:data_gen}.

\begin{algorithm}
\DontPrintSemicolon
\KwIn{Number of nodes $N_n$, Number of edges $N_e$}
\KwOut{Service Call Graph $\mathcal{G}(N_n, N_e)$}

Initialize $\mathcal{G}(V,E)$, where $V = \{1,2,\ldots,N_n\}, E = \phi$\;

\For{$i \in \{2,3,\ldots,N_n\}$}{
    $j \gets$ sample a node from $\{1,2,\ldots,i-1\}$ randomly\;
    $E \gets E \cup \{(i \rightarrow j)\}$\;
}
\For{$k \in \{N_n, \ldots, N_e\}$}{
    $i \gets$ sample a node from $\{2,3,\ldots,N_n\}$\;
    $j \gets$ sample a node from $\{1,2,\ldots,i-1\}$\;
    \uIf{$(i \rightarrow j) \notin E$}{
        $E \gets E \cup \{(i \rightarrow j)\}$\;
    }
}
\Return{$\mathcal{G}$}
\caption{Generate Random Graph}
\label{alg:generate_graph}
\end{algorithm}

\begin{algorithm}
\DontPrintSemicolon
\KwIn{Real Data $\mathcal{D}_{real}$, Service Call Graph $\mathcal{G}$}
\KwOut{Synthetic/Semi-Synthetic Data $\mathcal{D}$}

\For{each dependency edge $i$ in Figure \ref{fig:metric_graph}}{
    Generate quadratic/Learn function $f_i$\;
}
Learn function $f_0 : W^{agg} \rightarrow R$ \tcp*[l]{$R = $ \# instances}

$\mathcal{D}_{exog} \gets $ random distribution /$\mathcal{D}_{real}[W^{agg}]$\;

\For{$t = 1,2,\ldots$}{
\For{service $i$}{
    \uIf{$i$ = exogenous service}{
        $W^{i,agg}_{t} \gets \mathcal{D}_{exog}^{t}$\;
    }
    \Else{
        $W^{i,agg}_{t} \gets \sum_{j \in \mathcal{P}(i)} \beta_{j,i} * W^{j,agg}_{t}$\;
    }
    
    Compute $R_t$ using $f_0$\;
    For each instance $j$,  $W_{ijt} \gets N(\mu=\frac{W^{i,agg}_t}{R_t}, \sigma=\frac{\mu}{10})$\;
    
    Compute CPU and Mem. Util. of $i$ based on $f_i + \epsilon$\;
    
    \uIf{$i$ is leaf node}{
        compute Latency and Error based on $f_i+ \epsilon$\;
        Recurse back\;}
    
    \For{$ k \in $ child($i$)}{
        Goto Step 6 for service $k$\;
    }
    
    \uIf{all child($i$) is computed}{
        Compute $L^{agg}$ and $E^{agg}$ for each child($i$)\;
        Assign Latency and Error based on $f_i + \epsilon$\;
    }
}
}
$\mathcal{D} \gets$ all data generated for each service
\Return{$\mathcal{D}$}
\caption{Generate Synthetic/Semi-synthetic Data}
\label{alg:distributed_data}
\end{algorithm}

Algorithm \ref{alg:distributed_data} shows the steps required to generate synthetic or semi-synthetic data for a given call graph $\mathcal{G}$. We first create workload metric values for the exogenous nodes, that is, the nodes without any parent metrics. Workload for the service is distributed to its instances almost equally (Mean being total workload over number of instances). This is based on the observation from the real data where workload at instances were almost equal. Cpu and memory utilization values were computed based on the learned/generated functions, and then the metrics for the child services were computed similarly in a recursive nature. It is to be noted that we add a random gaussian error to the values of the metrics to avoid deterministic relationship of metric values across services.

\end{document}